\documentclass{aa-mod}
\usepackage{graphicx}
\usepackage{bm}
\pdfoutput=1

\usepackage{hyperref}
\usepackage{xcolor}
\usepackage{caption}
\usepackage{amsmath}

\begin{document}

\title{Characterization of Hamamatsu 64-channel TSV SiPMs}

\author{Max Renschler \inst{1} \fnmsep \thanks{\email{Max.Renschler@kit.edu}}
\and
William Painter\inst{1}
\and
Francesca Bisconti\inst{1}
\and
Andreas Haungs\inst{1}
\and
Thomas Huber\inst{1} \fnmsep \inst{2}
\and
Michael Karus\inst{1}
\and 
Harald Schieler\inst{1}
\and
Andreas Weindl\inst{1}
}

\institute{Karlsruhe Institute of Technology (KIT), Institut f\"ur Kernphysik, 76021 Karlsruhe, Germany
\and
DESY, 15735 Zeuthen, Germany
}

\abstract{}{
The Hamamatsu UV-light enhanced 64-channel SiPM array of the newest generation (\textit{S13361-3050AS-08}) 
has been examined for the purpose of being used for the Silicon Elementary Cell Add-on (SiECA) 
of the EUSO-SPB balloon experiment.
At a room temperature of $19.5\, \textrm{C}^{\circ}$, the average measured breakdown voltage of the array is 
$(51.65 \pm 0.11) \,$V, the average gain is measured to $(2.10 \pm 0.07) \cdot 10^6$ and the 
average photon detection efficiency 
results to $(44.58 \pm 1.80)\%$ at a wavelength of $(423 \pm 8)\, \textrm{nm}$ and a bias voltage of $55.2\,$V. 
The average dark-count rate is $(0.69 \pm 0.12)\,$MHz, 
equivalent to a dark-count rate per SiPM area of $(57 \pm 12)\, \textrm{kHz} / \textrm{mm}^2$, 
and the crosstalk probability is measured to $(3.96 \pm 0. 64)\%$. 
These results confirm the information given by the manufacturer.
Measurements performed with the newly installed Single Photon Calibration Stand at KIT (SPOCK)
show the improved sensitivity to photons with wavelengths lower than $400\,$nm 
compared to the SiPM array \textit{S12642-0808PA-50}, which was also investigated for comparison.
Additional measurements confirm the strong temperature dependence of 
the SiPM characteristics as given in the data sheet.
All the characterized parameters appear to be sufficiently uniform to 
build up a focal surface of SiPM arrays fulfilling the requirements 
for a telescope detecting photons in the UV range.\ \\  
}{}{}{}

\maketitle

\section{Introduction}
Silicon Photomultipliers (SiPMs) are a new type of photo detectors~\cite{Renker:2009zz} 
capable of 
detecting single photons with photon detection efficiencies in reasonable operation 
of up to 40\% and single photon time resolutions of several hundred picoseconds~\cite{timeResolution}. 
Due to the great progress of SiPM development during the recent years, SiPMs 
are in many ways comparable to conventional Photo Multiplier Tubes (PMTs) which have ruled the 
field of photon detection for decades. 
SiPMs have several advantages including a much lower operation voltage, a 
lightweight and robust structure and an insensitivity to magnetic fields \cite{SiPM_Char}. 
SiPMs are robust against too much incident light: Instead of burning the collecting 
anode of a PMT, a SiPM enters saturation and draws constant current but it is not damaged. 
However, SiPM properties are highly temperature dependent. The dark-count rate of SiPMs at room 
temperature is in the MHz regime which is much higher than the kHz dark-count rate of conventional 
PMTs.\\

The present task of characterizing SiPM arrays is embedded in the Extreme Universe Space Observatory 
(EUSO) project. EUSO aims to detect Extensive Air Showers (EAS) induced in 
Earth's atmosphere by ultra-high energy cosmic rays. Instead of detecting EAS by looking from the 
Earth up into the atmosphere, EUSO aims to detect EAS by looking from space down into the 
Earth's atmosphere~\cite{JEM-EUSO}. EAS can be detected by measuring fluorescence light 
emitted by nitrogen molecules excited during 
an EAS. Nitrogen emits fluorescence photons mostly in the UV range between $300\,-400\,$nm. 
The expected flux of fluorescence photons during an EAS measured from a 
height of around $400\,$km above ground, which resembles the flight height of the International 
Space Station (ISS), is around 400 photons per $2.5\,\mu$s and square meter for a $10^{20}\,$eV EAS \cite{EUSO-Exposure}. 
Looking at these limits, a good photon detection efficiency in the UV range and the 
ability of measuring single photons are constraints for the photon detection device used in EUSO. 
In the standard design, the focal surface of the EUSO telescope 
consists of 64-channel Multi-Anode PMTs (MAPMTs). \\
A technological evaluation of the EUSO detection system with a scaled down telescope was flown as 
the payload of the NASA 2017 Super Pressure Balloon (SPB) mission with the expected duration of at 
least 60 days at an altitude of $33 \pm 1\,$km. In parallel with the main camera ($6\, \textrm{x} 
\,6\,$ MAPMTs arranged in nine $2\, \textrm{x} \,2\,$ elementary cells), the Silicon Elementary Cell 
Add-on (SiECA) camera was mounted in the optical focal plane. It was centered on one edge of the MAPMT 
camera such that a cosmic ray event could pass through the field of view of both 
cameras~\cite{SPB-Flight}. This opportunity to evaluate MAPMT and SiPM detection systems under 
identical conditions has provided extensive experience and data for ongoing evaluation of the two 
methods of low photon intensity signal detection. Due to the 12-days flight time that was realized 
on this experimental delivery system, additional tests are in preparation to evaluate the SiECA 
camera while analysis of the SPB flight data regarding possible EAS event(s) and UV background 
measurement is ongoing~\cite{SPB-SiECA}.\\

\begin{figure}
	\centering
	\includegraphics[width = 0.49\linewidth]{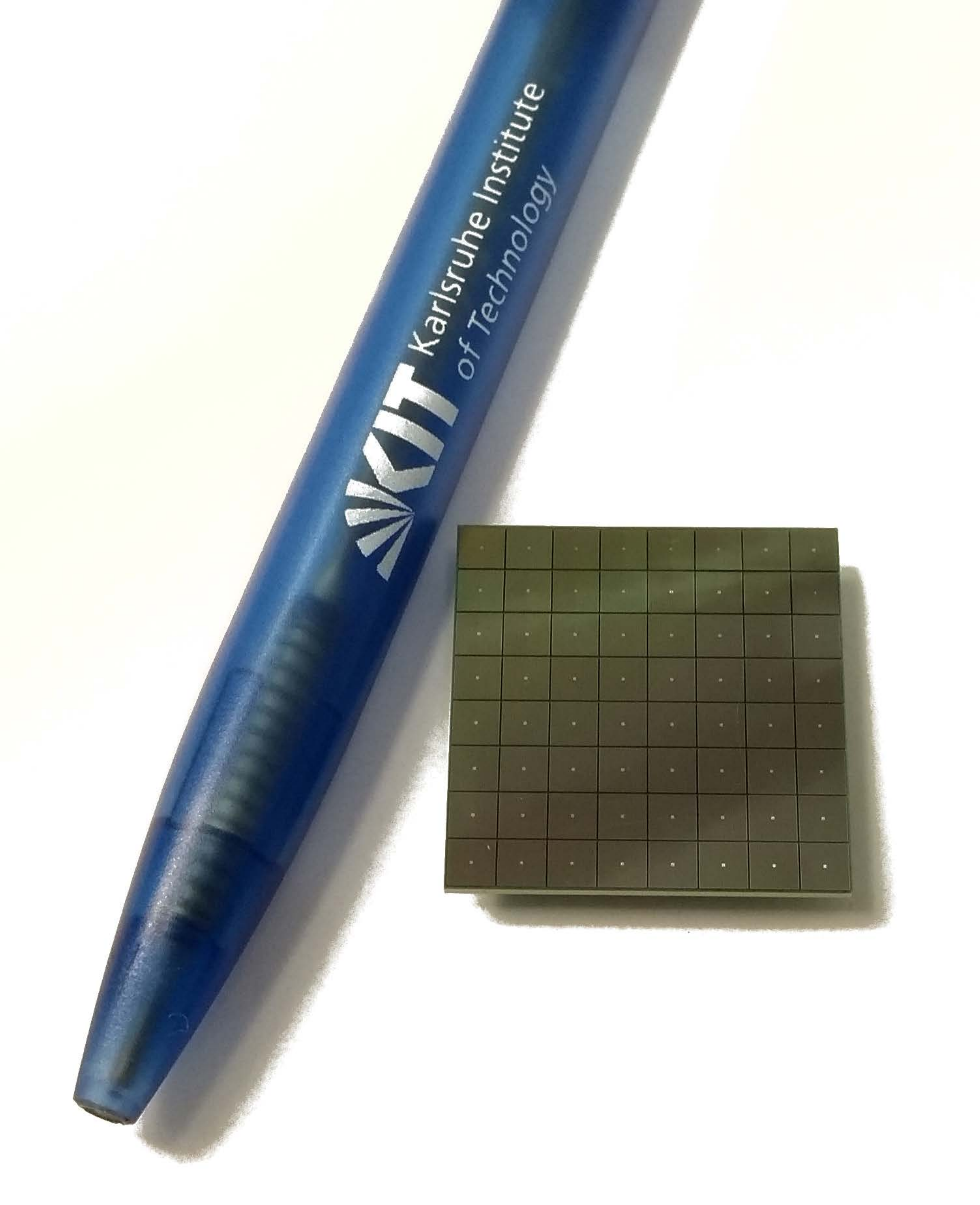}
    \includegraphics[width = 0.49\linewidth]{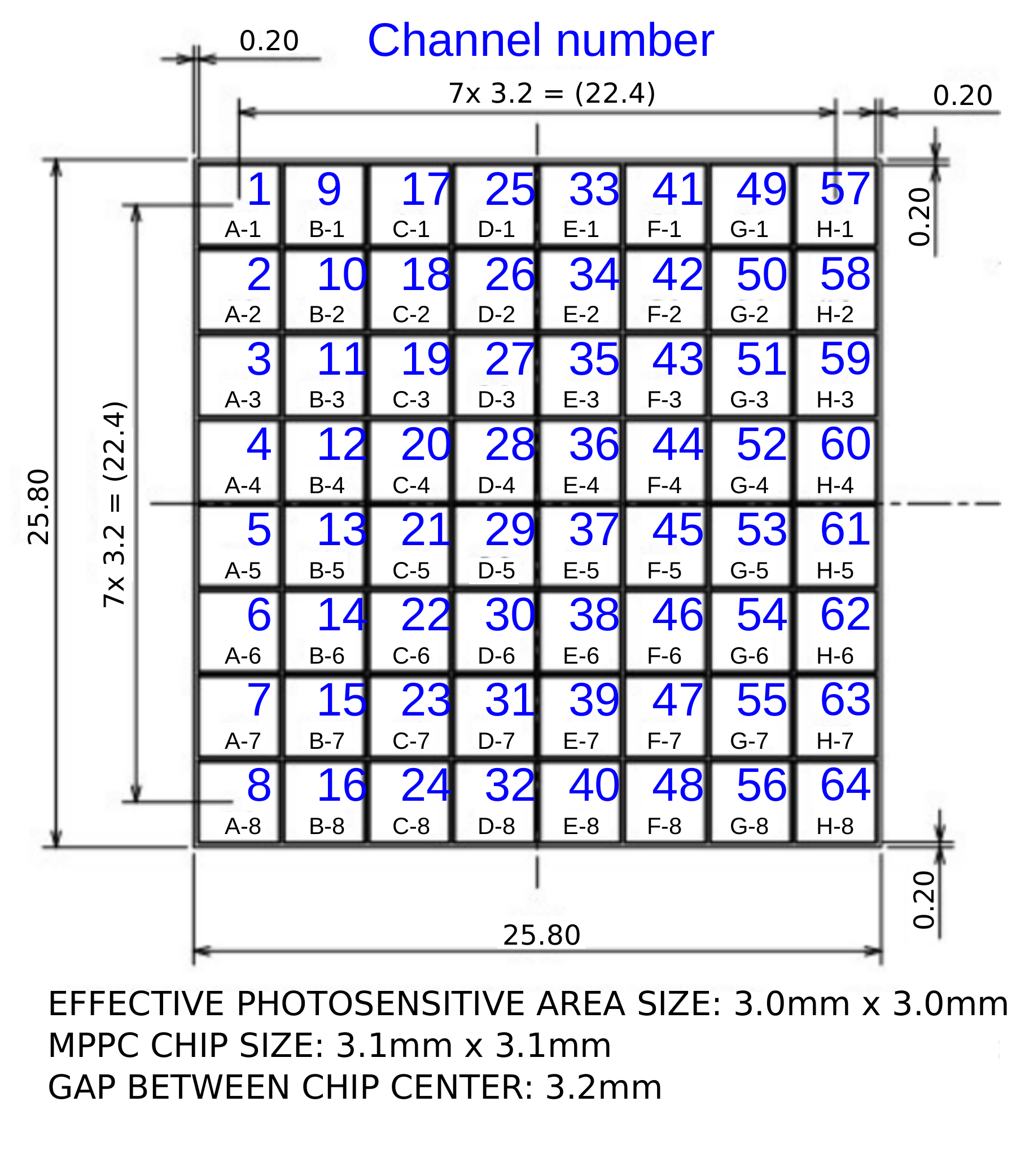}
	\caption{Left: 64-channel SiPM. Right: SiPM Layout (modified from \cite{S13Datasheet}).}
	\label{pic:SiPM_Array_Layout}
\end{figure}

Characterization results for the two 64 channel Hamamatsu SiPM arrays \textit{S12642-0808PA-50} 
(\textit{S12}) and \textit{S13361-3050AS-08} (\textit{S13}) are presented. 
The \textit{S12} SiPM array was released in 2014. 
The next series SiPM array \textit{S13} was released in 2015. 
Both SiPM arrays consist of 64 Through-Silicon-Via (TSV) SiPM channels with 3584 Avalanche 
Photo Diodes (APDs) in each channel and a pitch distance of $50\, \mu$m between the APDs. 
Figure \ref{pic:SiPM_Array_Layout} shows the ordering of the 64 channels and the size of the 
arrays which are the same for both arrays \textit{S12} and \textit{S13} (figure \ref{pic:SiPM_Array_Layout}). 
Through-Silicon-Via means that the cathode is etched through the silicon waver in the middle of 
the SiPM, building an electrical interconnection from the surface to the bottom of the SiPM device. 
Since the channel wires are attached on the back side of the SiPM array, the space between two 
channels can be reduced from about 3 mm without TSV to about $0.2\,$mm. 
Both arrays are equipped with \textit{Samtec ST4}\footnote{\textit{Samtec ST4-40-1.00-L-D-P-TR}} 
connectors on the backside.\\
An important difference between the two SiPM arrays is the coating of the SiPM array surface. 
The surface of the SiPM array \textit{S12} has an epoxy resin. The SiPM array \textit{S13} 
is coated with a silicone resin. The different resins result in a different sensitivity of 
UV photons with wavelengths less than $400\,$nm. With an epoxy resin, the sensitivity for light 
in the UV region goes down to a wavelength of $320\,$nm. With a silicone resin, light with a 
wavelength down to $270\,$nm can be detected. For more information about these specific SiPMs 
see~\cite{S13Datasheet,S12Datasheet}.\\
\section{Measurement setup: The Single Photon Calibration Stand at KIT} 

\begin{figure}
	\centering
	\includegraphics[width = \linewidth]{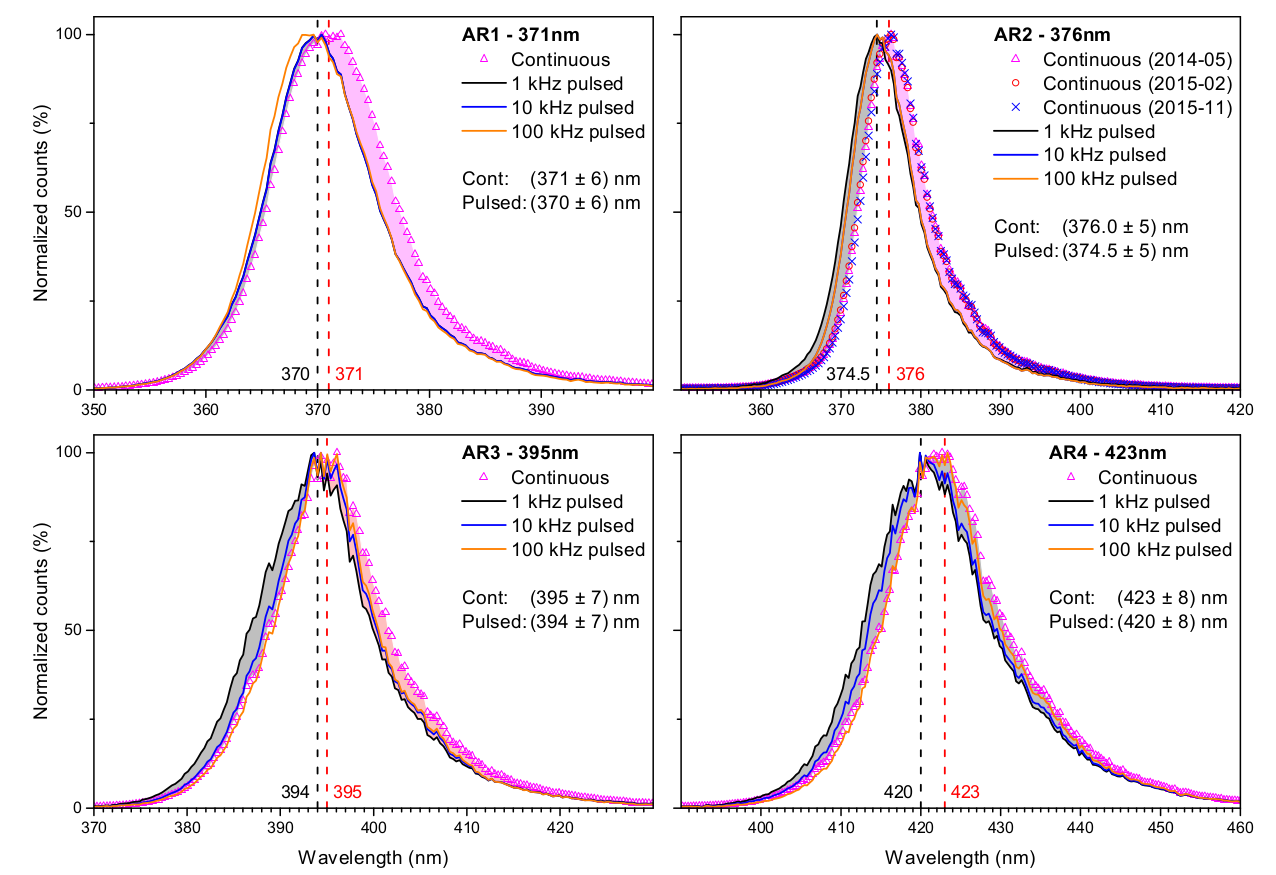}
	\caption{Measured continuous and pulsed spectra of the LED arrays~\cite{Michael}.}
	\label{pic:LED_spectra}
\end{figure}

\begin{figure}
	\centering
	\includegraphics[width = \linewidth]{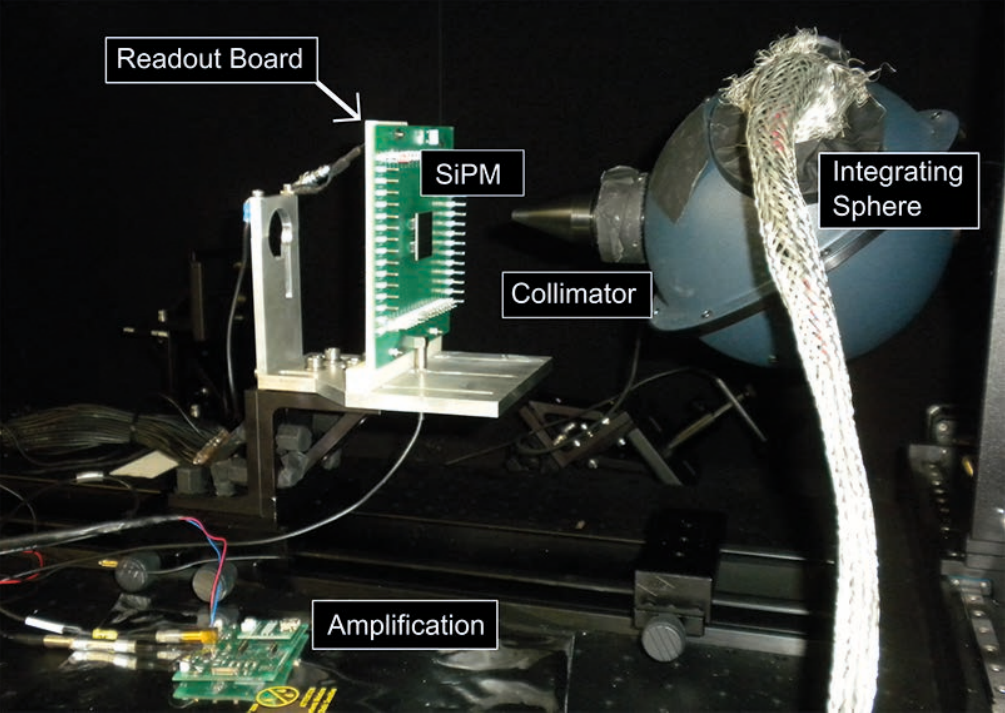}
	\caption{Picture of the SiPM measurement setup inside of the photon shielding of SPOCK.}
	\label{pic:SPOCK_inner_view}
\end{figure}

\begin{figure}
	\centering
	\includegraphics[width = \linewidth]{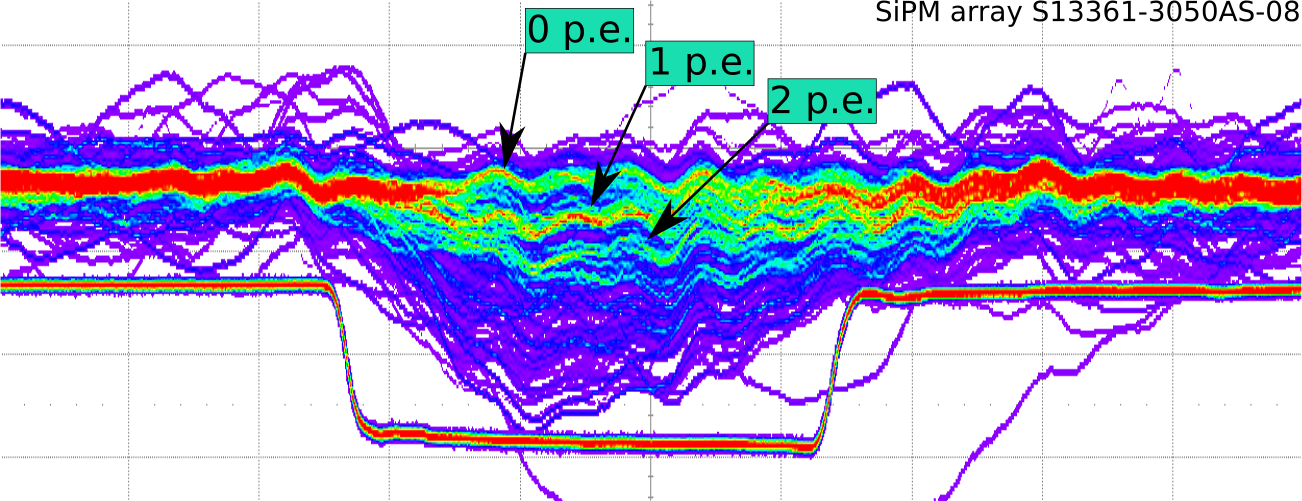}
	\caption{Persistence measurement of the SiPM signal and the gate time of $73\,$ns. The QDC starts integrating the signal $15\,$ns after receiving the gate signal. The pedestal and the first and second p.e. levels of the SiPM signal are clearly visible. The resolution of the SiPM signal is 10mV/div. The resolution of the gate signal is 500mV/div. The time resolution is 20ns/div.}
	\label{pic:SiPM_Gate}
\end{figure}

The measurements were performed with the Single Photon Calibration Stand at KIT (SPOCK). SPOCK consists of photon shielding, a reference light source, read-out electronics and measurement-controlling software. The photon shielding is a light-tight wooden box covered inside with black flock paper. The reference light source is placed inside of the photon shielding and consists of an integrating sphere, several LED arrays, a NIST-calibrated photo diode and a collimator. The integrating sphere is a \textit{Labsphere 3P-GPS-053-SL} with a diameter of $13.5\,$cm. Four LED arrays, each equipped with several LEDs of the same type and wavelength, are available. These can be attached to the entrance port of the integrating sphere. 
A single LED in the middle of the array can be used in pulsed mode (see table~\ref{tab:Setup-SPOCK_LEDarrays}). Measurements of the emitted wavelength of the LEDs were performed with an \textit{Ocean Optics S2000} spectrometer and the corresponding analogue-to-digital converter \textit{Ocean Optics ADC1000-USB} in pulsed and continuous mode (see \citep{Michael}). The results are shown in figure~\ref{pic:LED_spectra}. The characterization measurements presented here have been made with the LED array with wavelength $(423 \pm 8)\,$nm.\\
\begin{table*}
\centering
\begin{tabular}{cccc}
LED-array & LED & Number of LEDs & Wavelength (nm) \\
\hline \\
Array 1 & \textit{UVLED365-110E} & 42+1 & $371 \pm 6$\\
Array 2 & \textit{XSL-375-3E} & 20+1 & $376 \pm 5$\\
Array 3 & \textit{VL390-5-15} & 42+1 & $395 \pm 7$\\
Array 4 & \textit{VL425-5-15} & 12+1 & $423 \pm 8$\\
\end{tabular}
\caption{Overview of the available LED-arrays~\cite{Michael}. Shown are the used LEDs, the number of LEDs (continuous mode + pulsed mode) and the wavelength of the LEDs together with the statistical uncertainty for each LED array.}
\label{tab:Setup-SPOCK_LEDarrays}
\end{table*}
The NIST-calibrated photo diode measures the optical power inside of the integrating sphere. A collimator can be screwed to the exit port of the integrating sphere to reduce the optical power leaving the integrating sphere by a factor of $\approx 10^{-6}$. Together with the monitoring of the optical power inside of the integrating sphere, this allows emitted light with a known optical power onto the SiPM down to one photon per pulse in average. During the characterization measurements the optical power inside the sphere moved in the range of $180\,- 300\,$pW, resembling an output of two to four photons per pulse after the collimator.\\
The optical power during the measurement of each SiPM channel is monitored and stored. The overall change of the optical power results from variations in the output power of the light diode driver during the whole measurement time.\\
\begin{figure}
	\centering
	\includegraphics[width = \linewidth]{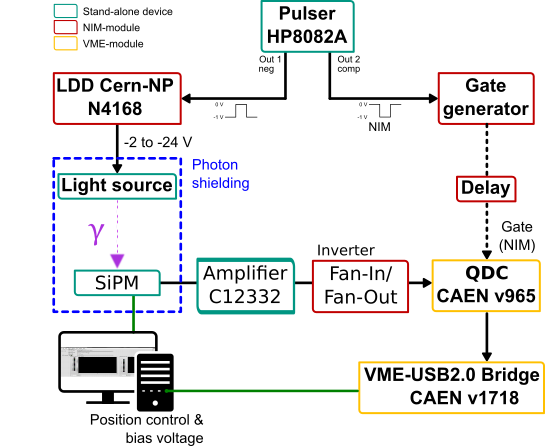}
	\caption{Flowchart of the measurement setup (modified from \cite{Michael}).}
	\label{pic:Setup-flowchart}
\end{figure}
A pulse generator \textit{HP8082A} generates simultaneously two pulses which are guided to a Light Diode Driver (LDD) and a gate generator (see figure ~\ref{pic:Setup-flowchart}). The LDD generates a variable pulse to drive the pulsed LED in the LED array. The gate generator generates the gate during which the Charge-to-Digital Converter (QDC) integrates the incoming SiPM signal.\\  
The SiPM is placed on a read-out board a few millimetres in front of the collimator exit. The board is a customized two-layer PCB\footnote{Printed Circuit Board} which guides the cathode signal of each SiPM channel to a pin connection. A channel is selected by closing the pin connection of the particular channel, i.e.\ only one channel can be read out at a time.\\
The SiPM signal is passed via a coaxial cable to the amplifier circuit of a SiPM Evaluation Board \textit{C12332} manufactured by Hamamatsu. The \textit{C12332}-Board has been modified for the purpose of amplifying incoming lemo signals. The on-board amplifier IC \textit{Texas Instruments OPA864} amplifies the SiPM signal by a factor of $(10 \pm 0.5)$. Figure~\ref{pic:SPOCK_inner_view} shows the measurement setup inside SPOCK.\\
The amplified SiPM signal is passed to a Fan-in/Fan-out which inverts the SiPM signal before it is guided into a QDC \textit{CAEN v965}. The QDC integrates the signal's charge during a variable gate time. The integration starts $15\,$ns after the gate signal has been received by the QDC and ends with the end of the gate signal. For the present measurements, the gate length was chosen to be $73\,$ns including the $15\,$ns activation time of the QDC. This gate length covers most of the SiPM signal and avoids probable after-pulses which should not be recorded (Fig.~\ref{pic:SiPM_Gate}). The gate is set with the help of delays between gate generator and QDC in a way that the QDC starts integrating the SiPM signal when signals induced by the pulsed LED are expected. The digitized integrated charge of the signal is recorded with the measurement software written in Labview.\\
The temperature is monitored during all measurements with a temperature sensor \textit{DS18B20} placed inside the shielding close to the SiPM. The measurement environment was cooled by an air condition. 
The temperature during the measurements was $(19.5 \pm 1)\,^{\circ}$C.\\
Temperature dependent measurements have been made to investigate the temperature dependency 
of the SiPM characteristics. For this, the SiPM was put into a $0.4\, \textrm{x} \,0.4\,$m$^2$ styrofoam cooling box together with thermal reservoirs, i.e.  water packs. The cooling box with the thermal reservoirs and the SiPM was cooled down in a fridge to $-10 ^{\circ}$C and measurements were made inside SPOCK as the system returned to ambient temperature.

\section{Measurement methods}
\begin{figure}
	\centering
	\includegraphics[width = \linewidth]{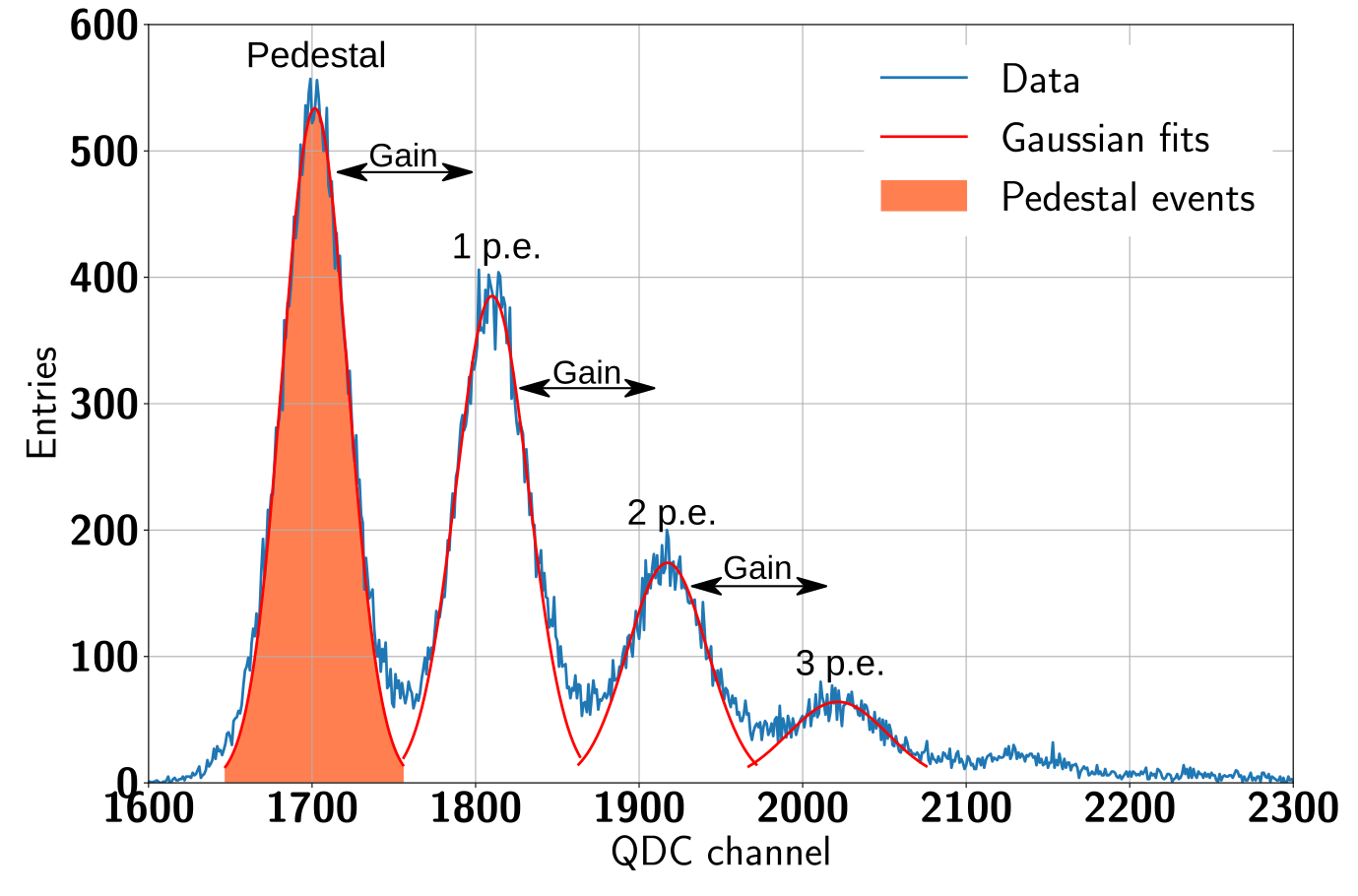}
	\caption{Example of a charge spectrum recorded with one channel of SiPM array \textit{S13}.}
	\label{pic:example_finger_spectrum}
\end{figure}

The measurement of the breakdown voltage, the gain, the photon detection efficiency (PDE), the dark-count rate and the crosstalk probability for each channel is made by recording a charge spectrum of the SiPM signals measured with the QDC. An example of a charge spectrum containing 70.000 QDC values is shown in figure~\ref{pic:example_finger_spectrum}. The pedestal peak consists of all the events in which no SiPM signal was recorded during a light pulse from the LED. The first photo electron (p.e.) peak contains all the events in which one APD fired and so on. Gaussian fits of the form
\begin{equation}
f(x) = \frac{A_{peak}}{\sqrt{2 \cdot \pi} \cdot \sigma} \cdot e^{-2 \cdot \left( \frac{x - x_{peak}}{2 \cdot \sigma} \right)^2}
\label{equ:SiPM-gaussian_fit}
\end{equation}
with $x$ in units of QDC channel, the peak position $x_{peak}$, the area under the Gaussian function $A_{peak}$ and the standard deviation $\sigma$ are performed to every individual peak in the charge spectrum.

The gain of a SiPM channel can be derived easily from the charge spectrum by measuring the distance between consecutive p.e. peaks. The Gain $G_{c}$ is measured in units of QDC channels. The real Gain $G$ can be calculated as
\begin{equation}
G = \frac{G_{c} \cdot k}{A \cdot e}.
\label{equ:SiPM-gain_transformation}
\end{equation}
with the QDC channel-to-charge transforming factor $k$, which is known from the calibration of the QDC~\cite{Michael}, the SiPM signal amplification factor $A$ and the elementary charge $e$. The amplification factor $A = 10 \pm 0.5$ has been measured by amplifying rectangular pulses of known height with variable widths up to some hundreds of nanoseconds. 

\begin{figure*}
\raggedright
\begin{minipage}{0.49\linewidth}
	\includegraphics[width = \linewidth ]{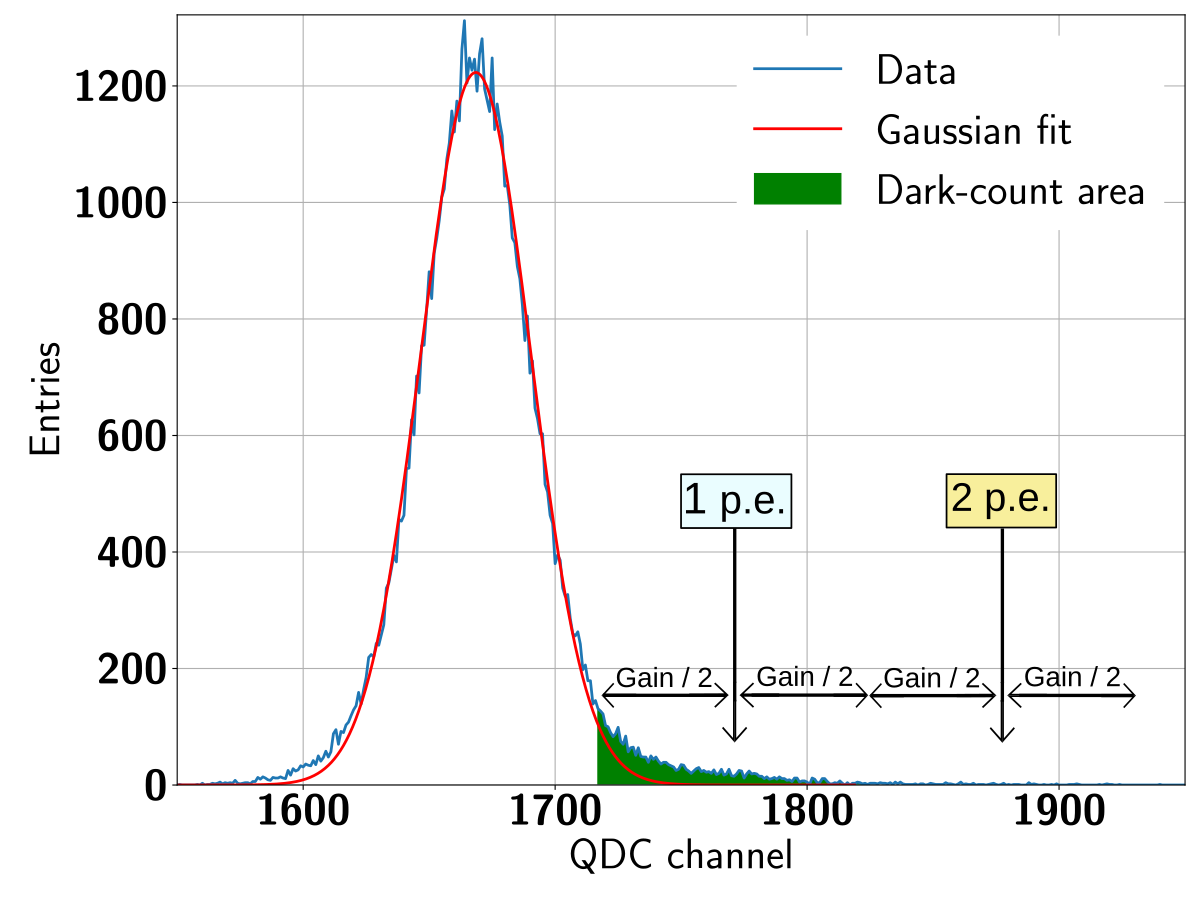}
\end{minipage}
\raggedleft
\begin{minipage}{0.49\linewidth}
	\includegraphics[width =\linewidth]{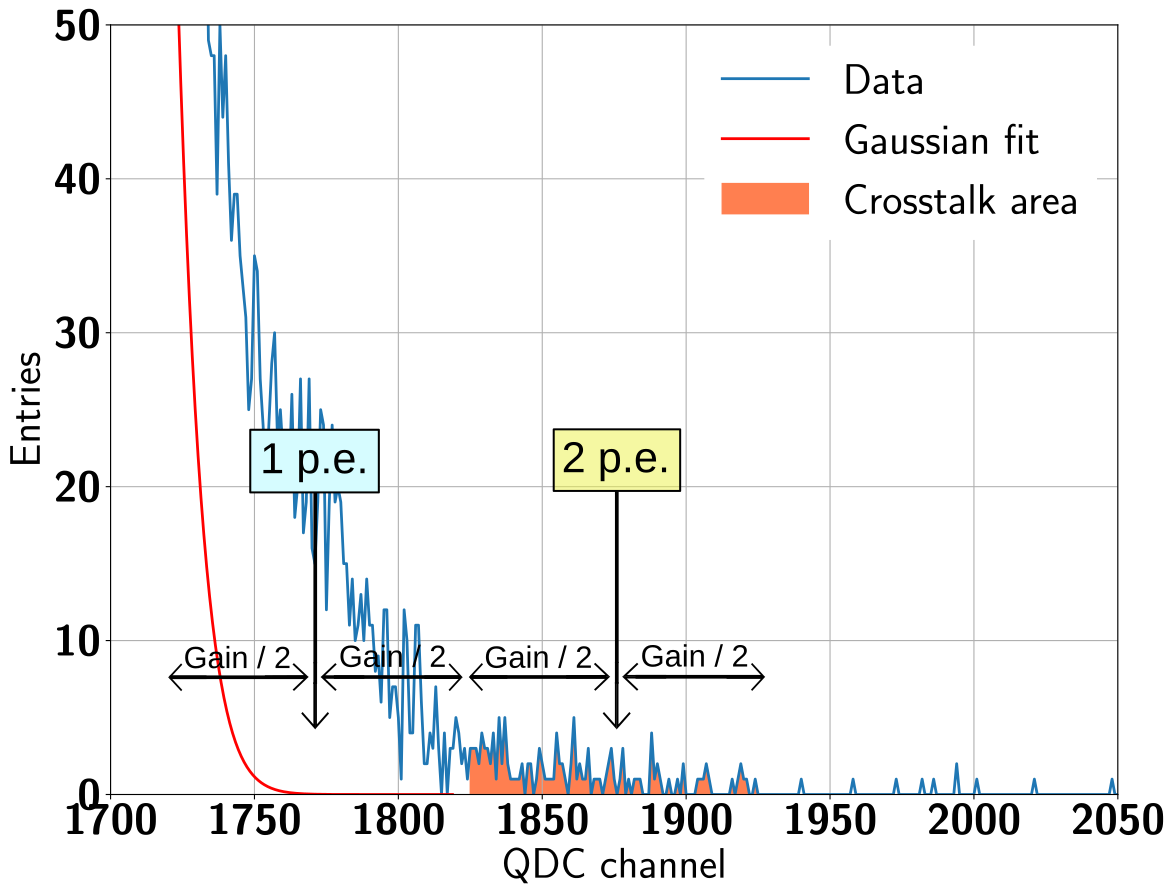}
\end{minipage}
\caption{Dark spectrum of a SiPM measurement. Left is the the dark spectrum with highlighted dark count area. Right is a zoom into the dark spectrum with highlighted crosstalk area. }
\label{pic:example_dark_spectrum}
\end{figure*}

The statistical and systematic uncertainties on the gain come from the statistical and systematic uncertainties on the measured distance between consecutive p.e.\ peaks $G_{c}$ and the channel-to-charge transforming factor $k$. The uncertainty of the amplification factor $A$ contributes only to the systematic uncertainty. 
\paragraph{Photon detection efficiency} The PDE can be calculated as
\begin{equation}
PDE = \frac{N_{pe}}{N}
\end{equation}
where $N_{pe}$ is the number of measured photoelectrons which equals the number of detected photons and $N$ is the number of incident photons per pulse. The number of incident photons is
\begin{equation}
N = \frac{P \cdot R \cdot R_{geom}}{E_{photon} \cdot f_{pulse}}
\label{equ:SiPM-N}
\end{equation}
where $P$ is the total optical power inside of the integrating sphere, $R$ is the collimator ratio, $R_{geom}$ is the geometrical fill factor which is the fraction of the incident light hitting the SiPM active area, $E_{photon}$ the photon energy of one photon and $f_{pulse}$ the light-pulse frequency. Since the collimator was placed right in front of the selected SiPM channel, the geometrical fill factor is assumed to be $R_{geom} = 1$.
For the number of the measured photoelectrons one can make use of Poisson statistics \cite{SiPM_Char}: The charge spectrum of an ideal SiPM has the form of a Poisson distribution 
\begin{equation}
P(k) = \left( \frac{\lambda^k}{k!} \right) \cdot e^{-\lambda}
\label{equ:Poisson}
\end{equation}
with the average number of detected photons per pulse $N_{pe}$ as mean value $\lambda$ and $k$ as the number of measured photoelectrons which equals the pedestal peak ($k = 0$, no photoelectron measured) and the p.e.\ peaks ($k = 1,2,..$) for an ideal SiPM. Due to crosstalk, afterpulsing and dark counts, the Poisson behavior of peaks after the pedestal is modulated. The pedestal itself is not influenced by crosstalk and afterpulsing and can be corrected for dark count events by calculating the dark count ratio with the help of a dark spectrum, i.e. a charge spectrum without incident light. So, the mean number of measured photoelectrons $N_{pe}$ can be derived by
\begin{equation}
N_{pe} = ln \left(\frac{N_{tot}}{N_{ped}}\right) - ln \left(\frac{N_{tot}^{dark}}{N_{ped}^{dark}}\right)
\label{equ:SiPM-Npe}
\end{equation}
where $N_{ped}$ and $N_{ped}^{dark}$ are the pedestal events (see figure \ref{pic:example_finger_spectrum}) in the measured charge spectrum  and the dark spectrum (see figure \ref{pic:example_dark_spectrum}). $N_{tot}$ and $N_{tot}^{dark}$ are the total numbers of events in the charge spectrum and the dark spectrum. 
The uncertainties of the PDE consist of the statistical and systematic uncertainties of the number of incident photons $N$ and the number of detected photons $N_{pe}$.
The statistical uncertainty of $N$ comes from the measurement of the light power inside the integrating sphere and the measurement of the collimator ratio. Regarding the systematic uncertainty, also the uncertainties of the LED wavelength and the pulse frequency come into play. 
The statistical uncertainty of $N_{pe}$ stems from the uncertainty of the number of pedestal events coming from the Gaussian fits to the pedestal peaks.
The systematic uncertainty of $N_{pe}$ comes from the systematic uncertainty in the number of pedestal events. This uncertainty is $\Delta_{N_{ped}} = \sqrt{2} \cdot 0.1\% \cdot N_{ped}$ with the number of measured pedestal events $N_{pe}$ and has its origin in the QDC non-linearity \cite{Michael}. \\
The average relative statistical and systematic uncertainties of $N$ and $N_{pe}$ are of more or less equal size in the region of 1\% (statistical uncertainty of $N$ and $N_{pe}$) to about 3.4\% (systematic uncertainty of $N$).
\paragraph{Breakdown Voltage} The breakdown voltage of a SiPM channel is determined by measuring the gain of the SiPM at different bias voltages above the assumed breakdown voltage. A linear regression is made regarding the gain-bias voltage behavior of the SiPM. The linear regression is extrapolated to the voltage value for which the gain of the SiPM becomes zero. This voltage is defined as the breakdown voltage of the SiPM channel. The gain was measured for 15 different bias voltages decreasing in voltage steps of $0.1\,$V. The linear regression of the gain-voltage dependence takes the statistical and systematic uncertainties of the gain measurement into account. The resulting uncertainty in the breakdown voltage comes from the uncertainties in the fit results and represents the sum of the statistical and systematic uncertainty.
\paragraph{Dark-count rate}  
To measure the dark-count rate, a dark spectrum of the SiPM is recorded with no light incident on the SiPM. This dark spectrum mainly consists of the pedestal peak (see figure \ref{pic:example_dark_spectrum}). Dark counts generate entries in the first p.e.\ peak. Due to possible crosstalk events induced by dark counts, also entries located in the second and maybe higher p.e.\ peaks have to be included. Although the p.e.\ peaks are not visible in the dark spectrum, the positions of the peaks can be found since all peaks are located at multiples of the gain with respect to the pedestal peak position. The peak area is assumed to be $\pm\, Gain/2$ around the p.e.\ peak position. To calculate the dark-count rate, the events in the area of the first and second p.e.\ peak are summed and divided by the total measurement time of the QDC. Dark count events higher than the second p.e.\ peak level are highly unlikely. To reduce the influence of noise to the measurement results, events located in higher p.e. peaks than the second p.e. peak are neglected.
\paragraph{Crosstalk probability} The crosstalk probability is determined with the same SiPM dark spectrum which is used for analyzing the dark-count rate. All pure dark count events are located around the first p.e. peak. If a dark count event induces a crosstalk photon which fires an adjacent APD, this will generate a second p.e. event. So all second p.e. events are considered as cross talk events (see figure \ref{pic:example_dark_spectrum}). Crosstalk events shared between adjacent SiPM channels can not be investigated with the actual measurement setup, but multi-channel crosstalk events are much less likely than single-channel crosstalk events due to the gap between adjacent channels.\\
The crosstalk probability is calculated by the fraction of the number of crosstalk events and the number of dark count events already determined. The statistical and systematic uncertainties of the number of dark count and crosstalk events are responsible for the uncertainties of the crosstalk probability.\\
\ \\
To determine the breakdown voltage, 15 charge spectra containing 15.000 QDC values each have been recorded. For measurements of the gain, the PDE, the dark-count rate and the crosstalk probability, charge and dark spectra with 70.000 QDC values have been used. Using these parameters, the characterization of one SiPM channel took about 20 minutes.

\section{Results}
\begin{figure}
\raggedright
\begin{minipage}{0.49\linewidth}
	\includegraphics[width = \linewidth]{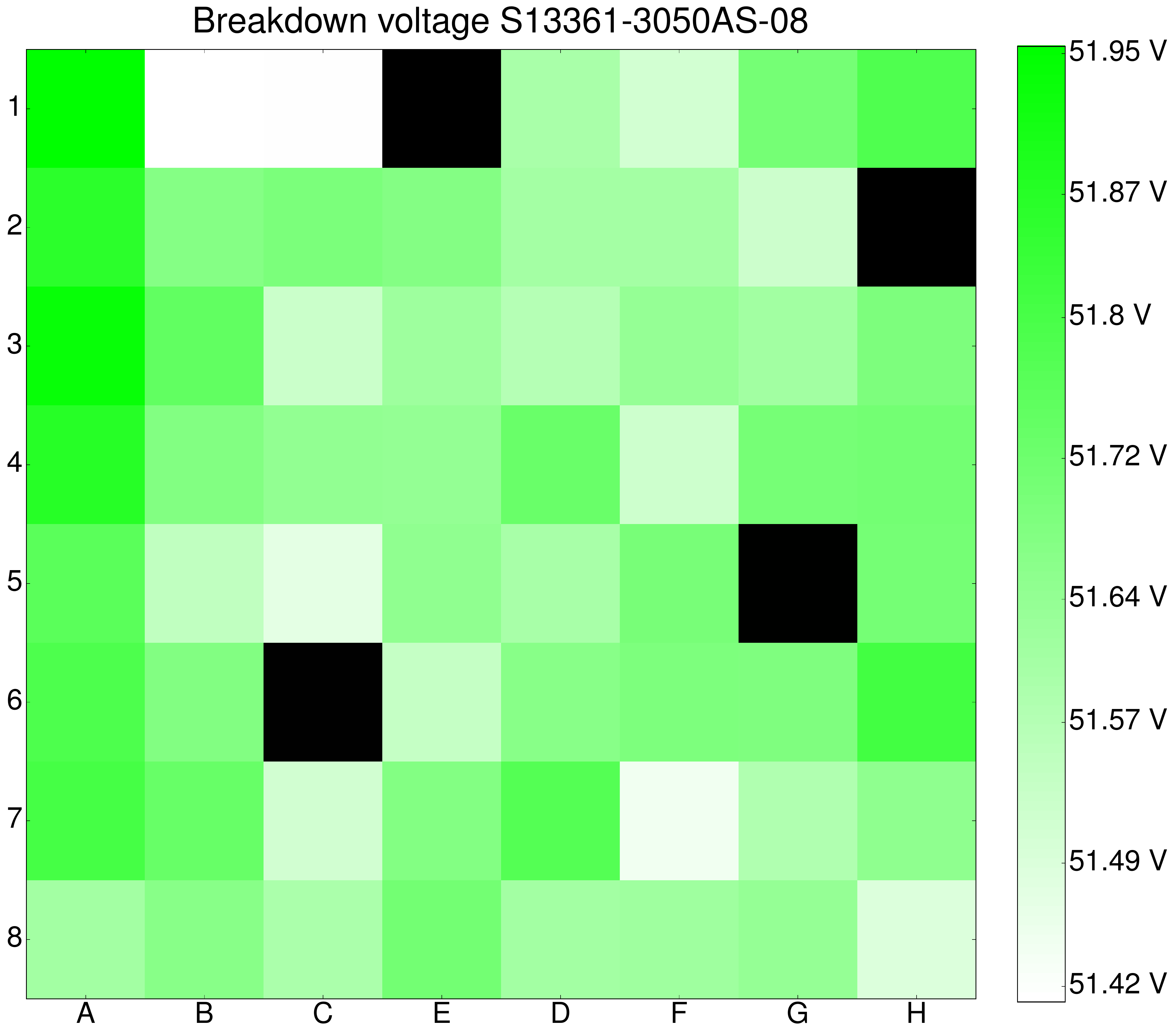}
\end{minipage}
\raggedleft
\begin{minipage}{0.49\linewidth}
	\includegraphics[width = \linewidth]{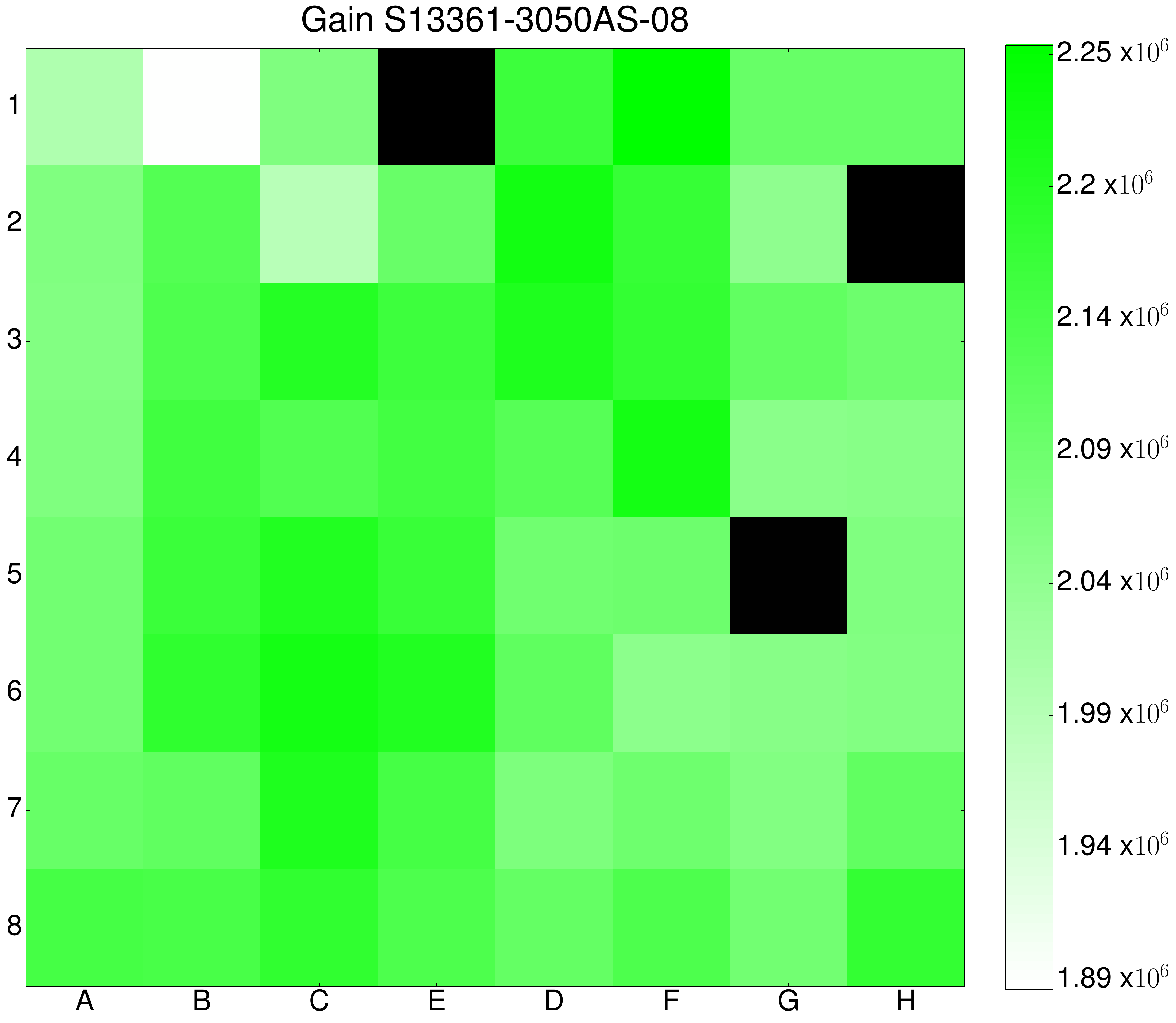}
\end{minipage}
\vspace{2mm}
\newline
\raggedright
\begin{minipage}{0.49\linewidth}
	\includegraphics[width = \linewidth]{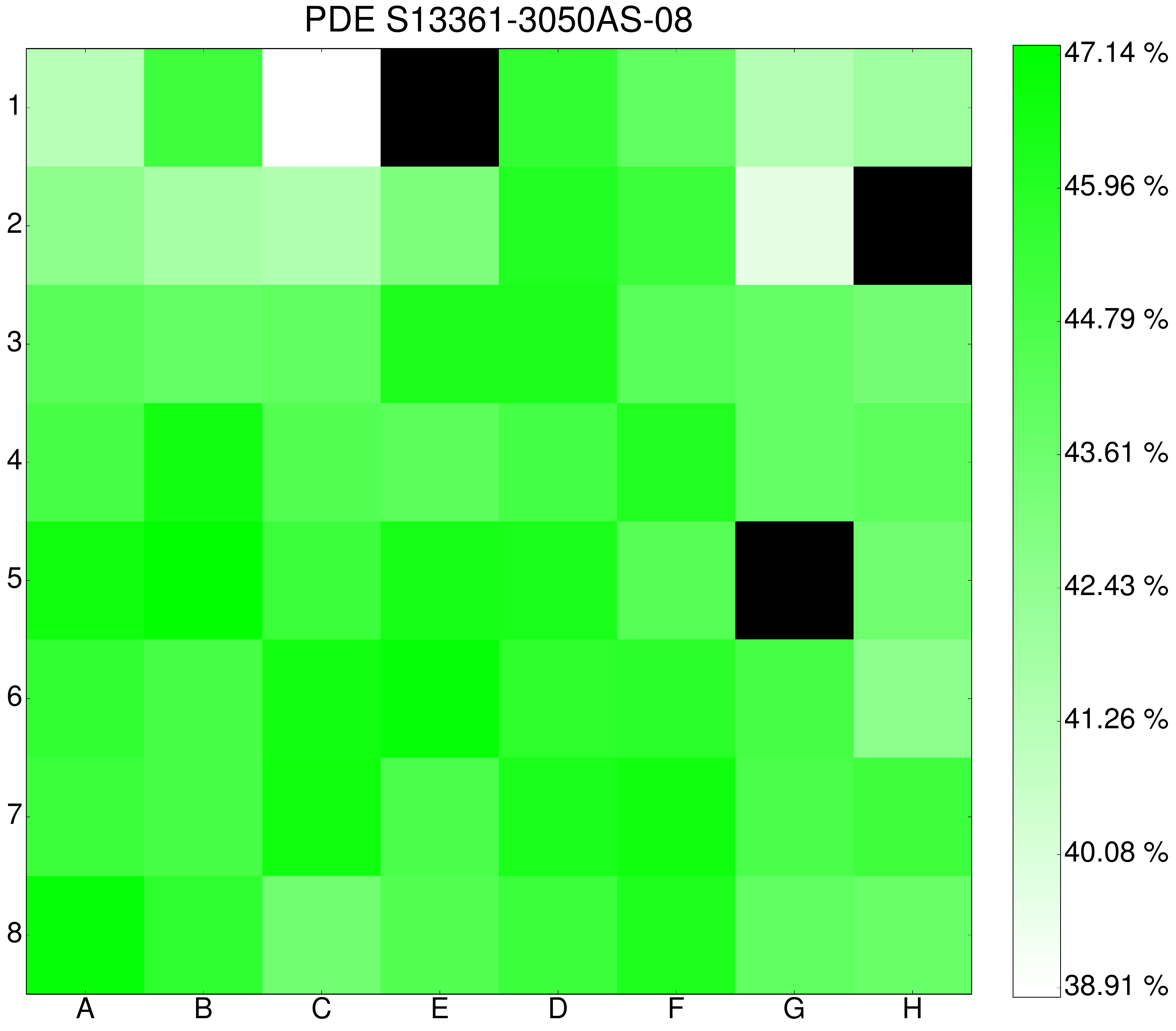}
\end{minipage}
\raggedleft
\begin{minipage}{0.49\linewidth}
	\includegraphics[width = \linewidth]{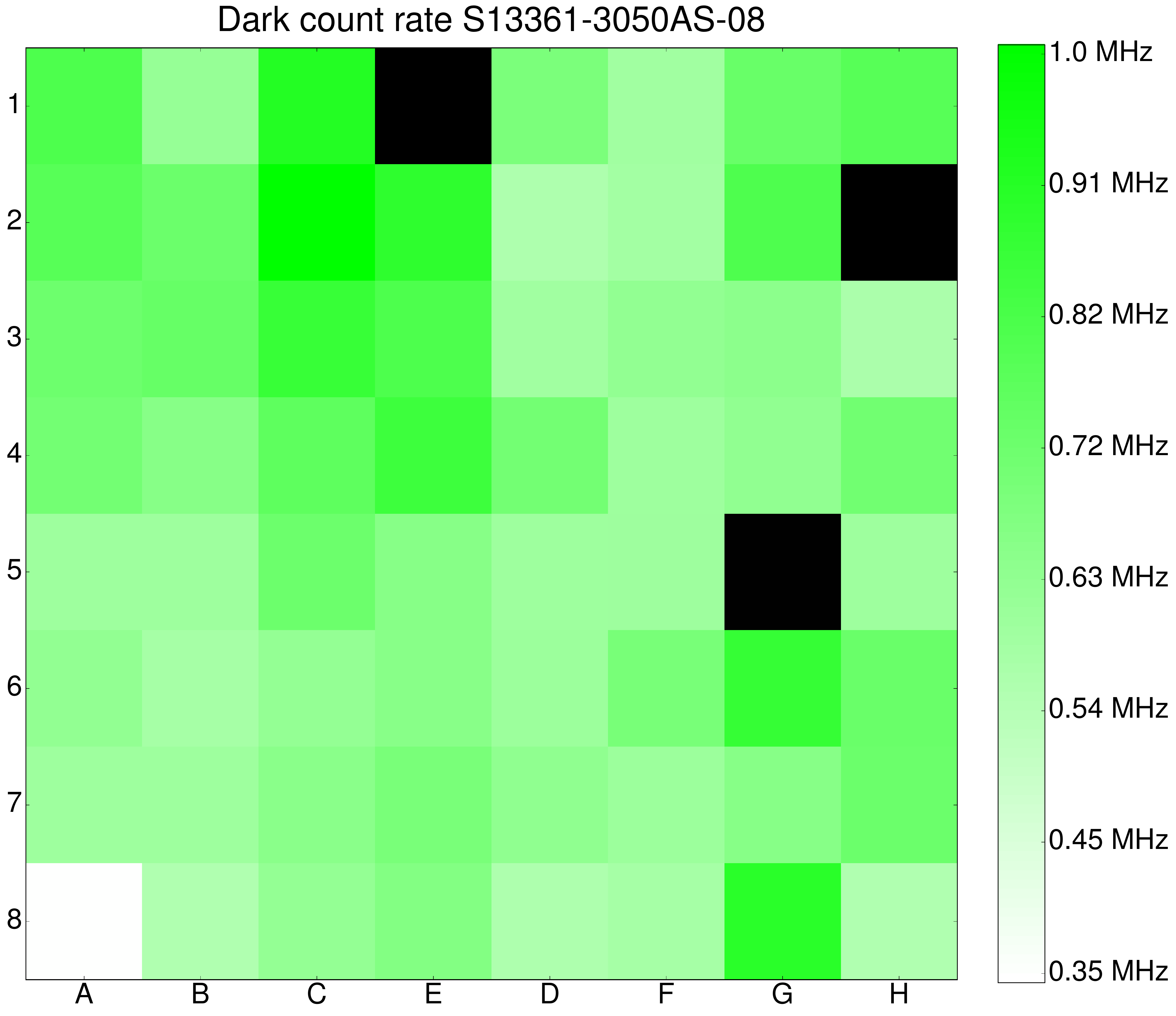}
\end{minipage}
\vspace{2mm}
\newline
\centering
\begin{minipage}{0.49\linewidth}
	\includegraphics[width = \linewidth]{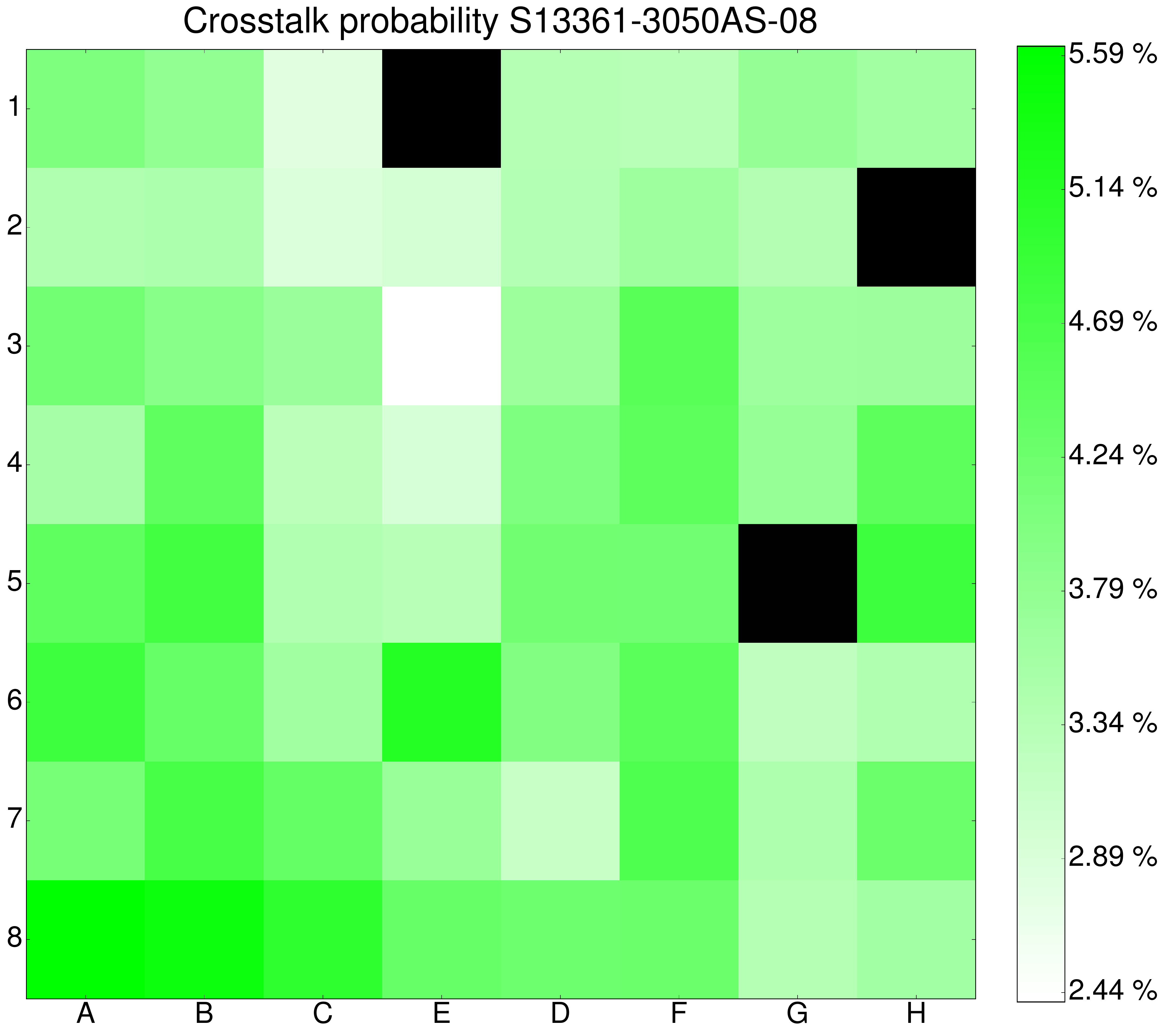}
\end{minipage}
\caption{Uniformity of the SiPM array \textit{S13361-3050AS-08}. Colour-coded heatmaps are shown of the breakdown voltage (up left), the gain (up right),
 the photon detection efficiency (middle left), the dark-count rate (middle right) and the crosstalk probability (bottom). Channels for which a measurement could not be performed are masked. }
 \label{pic:Heatmaps}
\end{figure}

One of the main results is that the SiPM array 
\textit{S13361-3050AS-08} has not only an improved UV-light sensitivity (see section~\ref{subsec:UV}), 
for which it was built, but also has better performance in most of the other properties measured. 
Therefore, the characterization results of the 64 channel SiPM array \textit{S13361-3050AS-08} 
(\textit{S13}) will be presented in detail. It is the newer product from Hamamatsu and finally 
chosen for SiECA. 
The characterization of the array \textit{S12642-0808PA-50} 
(\textit{S12}) from an older series will be briefly mentioned for comparison.\\
All the characterizing measurements of the SiPM array \textit{S13} have been made in a series 
at a constant bias voltage of $55.2\,$V and at a room temperature of $(19.5 \pm 1)\, \textrm{C}^{\circ}$. 
The incident pulsed light had a wavelength of $(423 \pm 8)\,$nm, a pulse frequency of $1\,$kHz 
and a pulse width of $50\,$ns. Unfortunately, the SiPM channels D1 (channel number 25, see fig. \ref{pic:SiPM_Array_Layout}), 
G5 (channel number 53) and H2 (channel number 58) could not be read out due to a wear out 
of the Samtec connectors on the read-out board. These channels are not used for the analysis.
\\
The results are given as the mean value of the results of all measured channels together with its standard deviation.
\begin{figure}
	\centering
	\includegraphics[width = \linewidth]{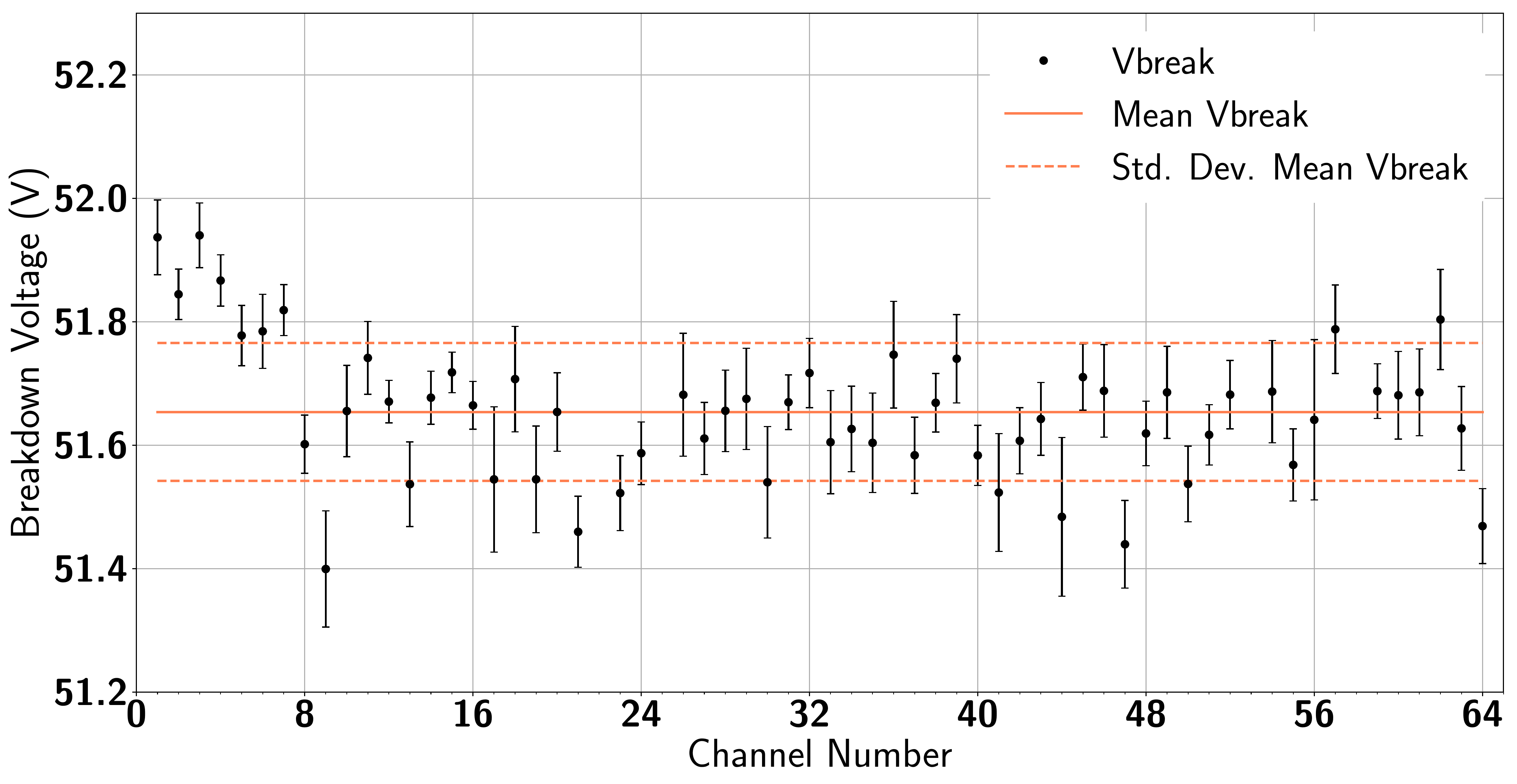}
	\caption{Breakdown voltage measurement of SiPM array \textit{S13361-3050AS-08}. The ambient temperature during the measurement was $(19.5 \pm 1)\,^{\circ}$C. The uncertainty shown stems from the fit results.}
	\label{pic:S13_Vbreak}
\end{figure}
\paragraph{Breakdown voltage}  
Figure~\ref{pic:S13_Vbreak} shows the result of the breakdown voltage measurement of the available SiPM channels. Unfortunately, the breakdown voltage measurement failed for channel C6 (channel number 22, see fig. \ref{pic:SiPM_Array_Layout}). Reasons for this might be a sudden electric noise pollution which was observed several times during the measurements or a temporary malfunction of the Samtec connectors. The mean breakdown voltage $V_{break}$ over the measured channels of the array is
\begin{equation}
V_{break} = (51.65 \pm 0.11)\,\textrm{V} 
\end{equation}
at a mean measurement temperature of $19.5^{\circ}$C. This value is consistent with the manufacturer's information of $(53 \pm 5)\, $V at $25^{\circ}$C. All breakdown voltage values are placed in a voltage band with a width of $0.5\,$V. 
Regarding a uniform operation for all SiPM channels in the array, these small inhomogeneities can easily be handled by modern ASICs for SiPM read-out if one power supply for each channel is not an option. In this case, a few power supplies generate the overvoltage and the ASIC fixes the slight differences between the channels due to the different breakdown voltages. This method is applied in SiECA where the ASIC \textit{Citiroc} manufactured by \textit{Weeroc} is used for this task~\cite{citiroc}. 
\begin{figure}
	\centering
	\includegraphics[width = \linewidth]{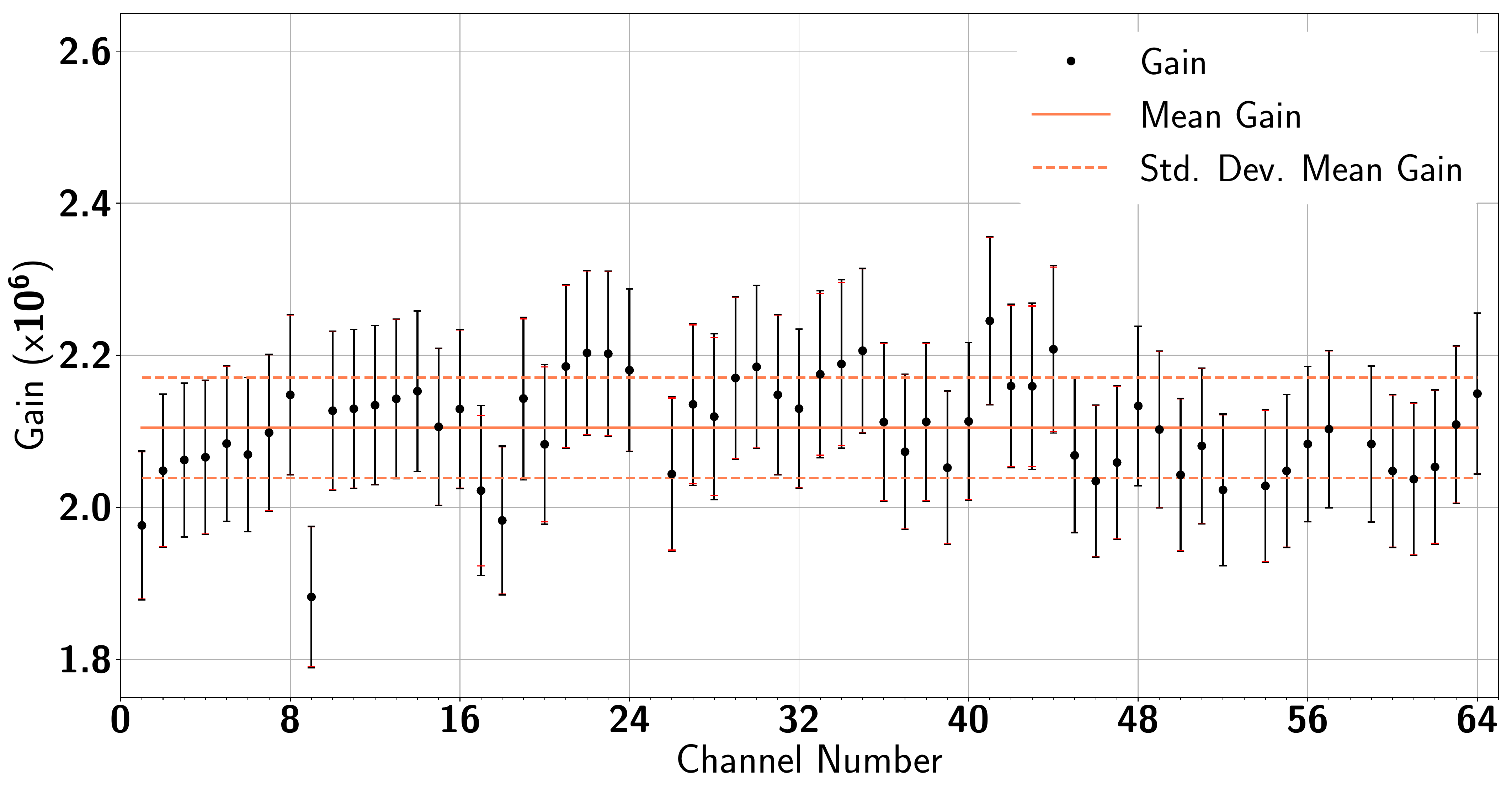}
	\caption{Gain measurement of SiPM array \textit{S13361-3050AS-08}. The bias voltage was $55.2\,$V and the incident light had a wavelength of $(423 \pm 8)\,$nm. The ambient temperature during the measurement was $(19.5 \pm 1)\, ^{\circ}$C. Statistical errors are shown in black, systematic errors are shown in red.
For this measurement, the statistical and systematic uncertainties are more or less of equal size. The red systematic errorbars are hidden behind the black statistical errors. 
}
	\label{pic:S13_Gain}
\end{figure}
\paragraph{Gain} The results of the gain measurement of array \textit{S13} are shown in 
figure~\ref{pic:S13_Gain}. The gain describes the amplification of the first photo electron 
generated by an incident photon. The mean gain $G$ over the whole array is
\begin{equation}
G = (2.10 \pm 0.07) \cdot 10^6 .
\end{equation}
The data sheet of \textit{S13} reports a gain of $1.7 \cdot 10^6$ at a 
temperature of $25\,^{\circ}$C, an overvoltage 
of $3\,$V and at a wavelength of $450\,$nm. Since the measurements presented here have been made at 
a higher overvoltage of $3.55\,$V, a slightly higher gain is expected. The inhomogeneity in the 
gain over the whole array can mostly be explained by the inhomogeneity of the breakdown voltage of 
the different channels of the array (see also fig.~\ref{pic:Heatmaps}). 
Since the characterization measurements were made with a constant bias voltage of $55.2\,$V, a lower 
breakdown voltage of a SiPM channel leads to a higher overvoltage during the characterization 
measurement with respect to the mean value. Since the gain depends strongly on the overvoltage 
of the SiPM, the measured gain of this channel will be higher than the mean gain value. 
\begin{figure}
	\centering
	\includegraphics[width = \linewidth]{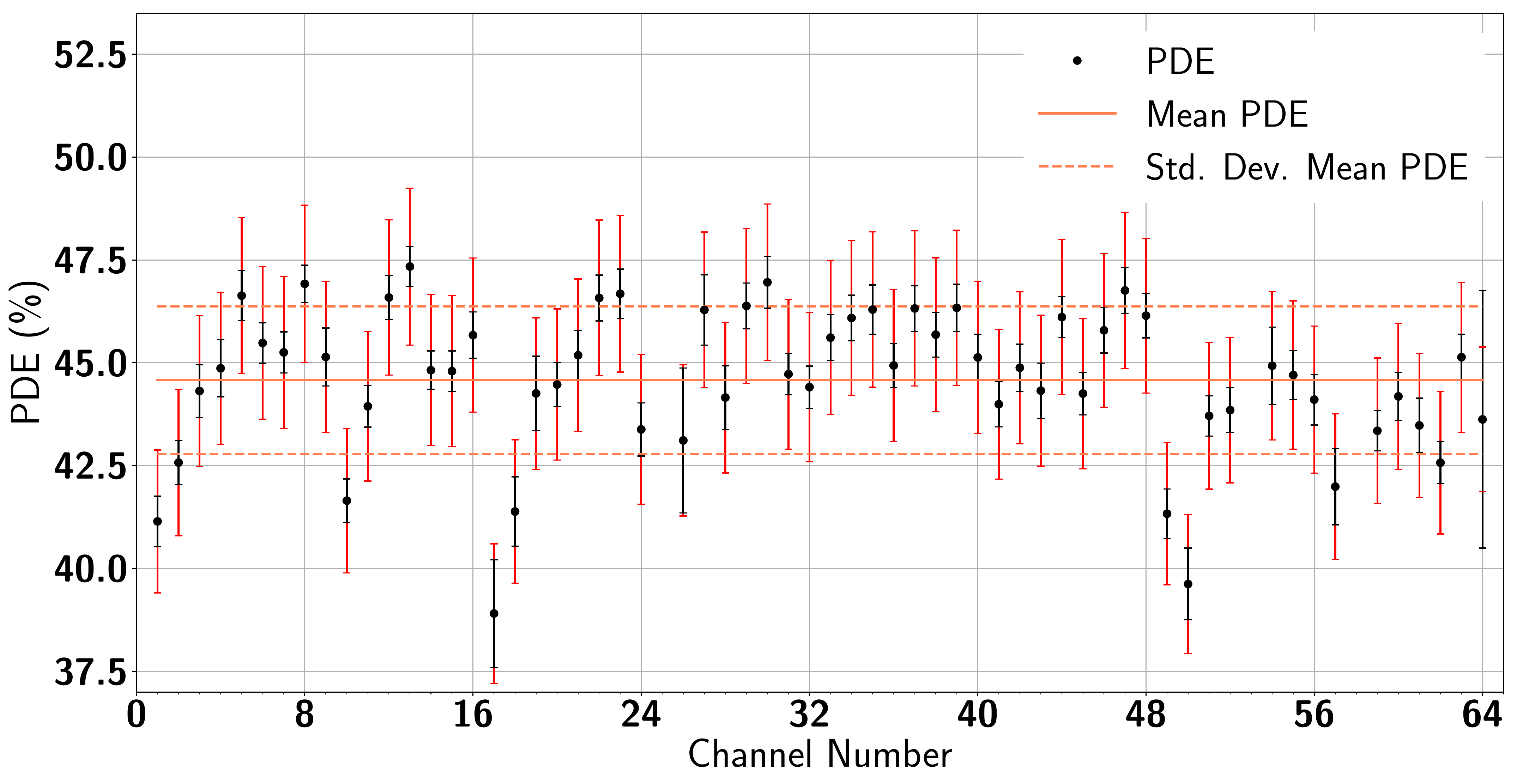}
	\caption{PDE measurement of SiPM array \textit{S13361-3050AS-08}. The bias voltage was $55.2\,$V 
	and the incident light had a wavelength of $(423 \pm 8)\,$nm. The ambient temperature during 
	the measurement was $(19.5 \pm 1)\, ^{\circ}$C. Statistical errors are shown in black, 
	systematic errors are shown in red.} 
	\label{pic:S13_PDE}
\end{figure}
\paragraph{PDE} In figure \ref{pic:S13_PDE} the results of the PDE measurement are shown. The mean PDE over the whole array is
\begin{equation}
PDE = (44.58 \pm 1.80)\,\%
\end{equation}
at a temperature of $19.5^{\circ}$C and an overvoltage of $3.55\,$V. This is about $5\,\%$ higher than the manufacturers prediction of $40\,\%$ which again can most probably be explained with the higher overvoltage used in the characterisation measurements. The measured result is comparable to a measurement of a one channel SiPM \textit{Hamamatsu S13360-3050CS} in~\cite{SiPM_Char2}.\\
\begin{figure}
	\centering
	\includegraphics[width = \linewidth]{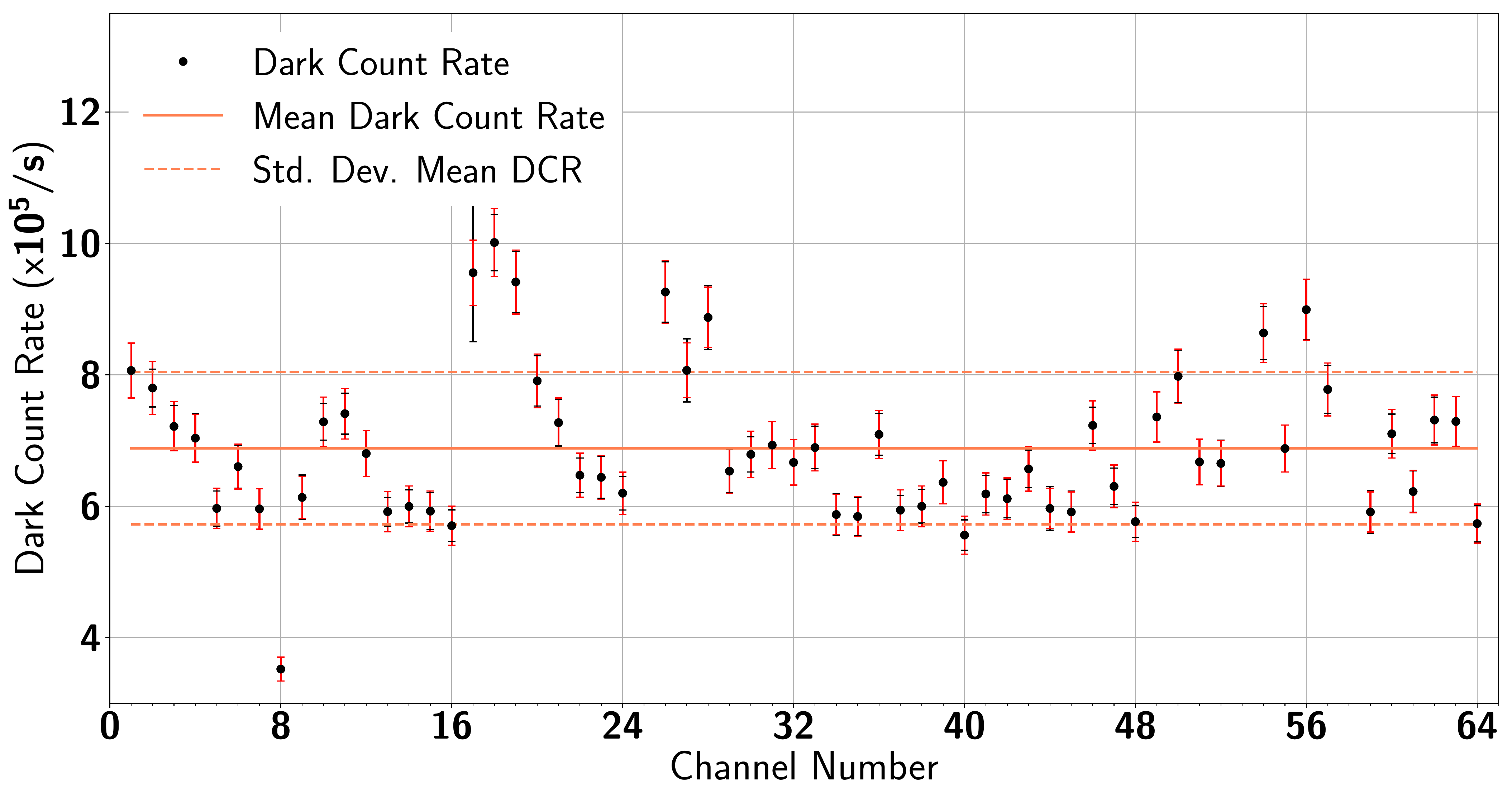}
	\caption{Dark-count rate measurement of SiPM array \textit{S13361-3050AS-08}. The bias voltage was $55.2\,$V. The ambient temperature during the measurement was $(19.5 \pm 1)\,^{\circ}$C. Statistical errors are shown in black, systematic errors are shown in red.}
	\label{pic:S13_DC}
\end{figure}
\paragraph{Dark-count rate}
The dark-count rate results are presented in figure~\ref{pic:S13_DC}. The mean dark-count rate per channel $\xi$ is
\begin{equation}
\xi = (6.88 \pm 1.16) \cdot 10^5 \,\textrm{Hz}.
\end{equation}
This is comparable to the dark-count rate given in the data sheet with $0.5$ to $1.5\,$MHz 
at $25^{\circ}$C and with an applied overvoltage of $3\,$V. Nevertheless, the measured variation 
in the dark-count rate of circa $16\%$ between the different channels of the array is apparently large. 
A reason for this might be the variation in the breakdown voltages between the channels. 
Also, since the dark-count rate is a highly temperature dependent characteristic 
(see figure \ref{pic:temp_DC}), the temperature change in the range of $\pm1^{\circ}$C during the
 measurement of different channels is expected to have a visible influence on the dark-count rate results (see section \ref{sec:temp} for more information).
\begin{figure}
	\centering
	\includegraphics[width = \linewidth]{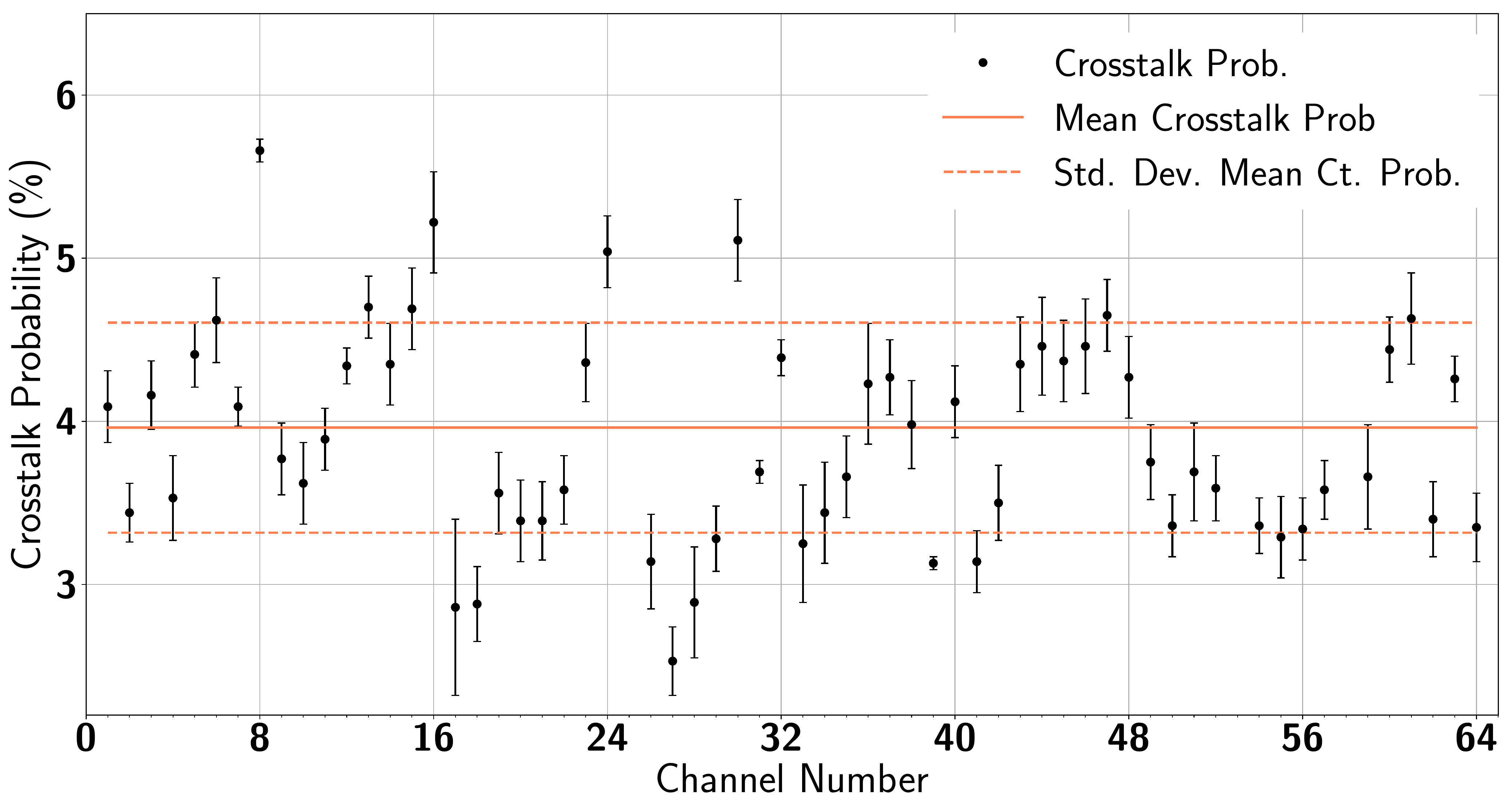}
	\caption{Crosstalk measurement of SiPM array \textit{S13361-3050AS-08}. The bias voltage was $55.2\,$V. The ambient temperature during the measurement was $(19.5 \pm 1)\, ^{\circ}$C. Statistical errors are shown in black. The systematic errors are plotted in red but are too small to be visible.}
	\label{pic:S13_CT}
\end{figure}
\paragraph{Crosstalk probability} 
The crosstalk probability is shown in figure \ref{pic:S13_CT}. The mean crosstalk probability $\epsilon$ is
\begin{equation}
\epsilon = (3.96 \pm 0.64)\,\%
\end{equation}
This value is less than half of the value of former SiPM generations for which crosstalk probabilities of more than $10\,\%$ are typical. The manufacturer claims a probability of $3\,\%$ at an overvoltage of $3\,$V. The distribution of the measured values over the array can be seen in figure~\ref{pic:Heatmaps}. Since the crosstalk probability depends on the number of adjacent APD cells, a lower crosstalk probability for channels at the edge of the array might be reasonable. Nevertheless, a lower crosstalk probability for channels at the edge of the array is not visible. A logical reason for this is the gap between the channels in the array of about $0.2\,$mm. Since this gap surrounds every channel and suppresses the exchange of crosstalk photons between different channels, the location of a channel in the array has no influence on the crosstalk probability.\\
\ \\

The characterizing measurements of array \textit{S12642-0808PA-50} were made at a bias 
voltage of $67.6\,$V with the same light source. Table~\ref{tab:measurements-comp.results} 
shows the measurement results of array \textit{S12} and \textit{S13} together with a comparison. 
The new series SiPM array \textit{S13} has improvements in all of the measured characteristics.\\
In addition to the channels D1 (channel number 25, see fig. \ref{pic:SiPM_Array_Layout}) , G5 (channel number 53) and H2 (channel number 58) 
which already could not be read out during the measurement of SiPM array \textit{S13}, 
the channels G4 (channel number 52) and G2 (channel number 50) could not be read out in this 
consecutive measurement of SiPM array \textit{12642} due to a wear out of the Samtec connectors 
on the read-out board.\\

Information about the uniformity of the measured characteristics of \textit{S13} are visualized in 
figure~\ref{pic:Heatmaps}. An overall uniformity of the single channels of the array is given and 
sufficient for building a homogeneous SiPM focal surface. Differences, especially in the single 
channel breakdown voltages, are small enough to be fixed by ASICs.\\
Nevertheless, the variations in, for example, the PDE or the dark-count rate seem quite large. 
By adjusting the bias voltage individually for every single channel of the array so that 
every channel works with the same overvoltage, the overall uniformity should increase 
significantly, which is subject of further studies using the next version of SiECA. 
For SiECA, the individual bias voltage adjustment has successfully been realized with the Citiroc ASIC 
to guarantee a homogeneous focal surface.

\begin{table*}
\centering
\begin{tabular}{cccc}
 & S12642 & S13361 & Comparison\\ 
\hline \\
Breakdown voltage (V) & $64.62 \pm 0.10$ & $51.65 \pm 0.12$ & -20 \% \\
Gain ($10^6$) & $1.65 \pm 0.04$ & $2.12 \pm 0.07$ & +28 \% \\
PDE ( \% ) & $35.69 \pm 1.09$ & $44.6 \pm 1.78$ & +25 \% \\
Dark-count rate (MHz) & $1.29 \pm 0.14$ & $0.68 \pm 0.11$ & -43 \% \\
Crosstalk probability (\%) & $11.17 \pm 1.27$ & $3.9 \pm 0.66$ & -65 \% \\
\end{tabular}
\caption{Summary of the average characterization results for the SiPM arrays \textit{S12642-0808PA-50} 
and \textit{S13361-3050AS-08}. Column three shows the improvements of the new series 
SiPM array \textit{S13} compared to the SiPM array \textit{S12}. 
For all measurements, light with a wavelength of $423\,$nm was used.}
\label{tab:measurements-comp.results}
\end{table*}

\subsection{Response behaviour}
In addition to the characterization measurements, the response behavior of the two SiPM arrays 
has been investigated. 
Due to the working principle of a SiPM, the signal is not always proportional to the number of 
incident photons. The reason for this is that the signal of the single Avalanche Photo Diode (APD) 
is not sensitive to the number of photons hitting the APD, but digitally fires or not. 
For large numbers of incident photons, the total SiPM signal saturates with available APDs activated. 
The SiPM response behaviour describes the SiPM signal regarding a variable number of incident photons. 
This is the number of firing APDs $N_{APD}$
with respect to the number of incident photons $N$  (see eq.~\ref{equ:SiPM-N}). 
For the response behaviour measurement, the same light source as for the characterization measurements 
with photons of a wavelength of $(423 \pm 8)\, $nm was used.\\
The number of firing APDs was determined by fitting a Poisson-like distribution of the form

\begin{equation}
P(k) = \left( \frac{\lambda^k}{k!} \right) \cdot e^{-\lambda} \cdot H
\end{equation}

to each measured charge spectrum. Here, $\lambda$ is the mean number of fired APDs and $k$ is the p.e. peak index (see equ. \ref{equ:Poisson}). To obtain reasonable fit results, the factor $H$ was included in the fit function to take care of the height of the charge spectrum. As data points for the Poisson fit, the maxima of the Gaussian fits of the single p.e. peaks have been chosen. To transform the position of the Gaussian maxima from QDC channel to p.e. peak index $k$, the position was divided by the gain measured with the actual charge spectrum. The overall spectrum was subtracted by the pedestal peak position to start with a value of $k = 0$ for the pedestal peak. If a gain measurement failed since the incoming light had a too low intensity to create a sufficient charge spectrum, the gain value measured during the characterization of the actual channel was chosen and the fit data points have been selected as the charge spectrum entries at the position of multiples of this gain.\\
\begin{figure}
\raggedright
\begin{minipage}{0.49\linewidth}
	\includegraphics[width = \linewidth ]{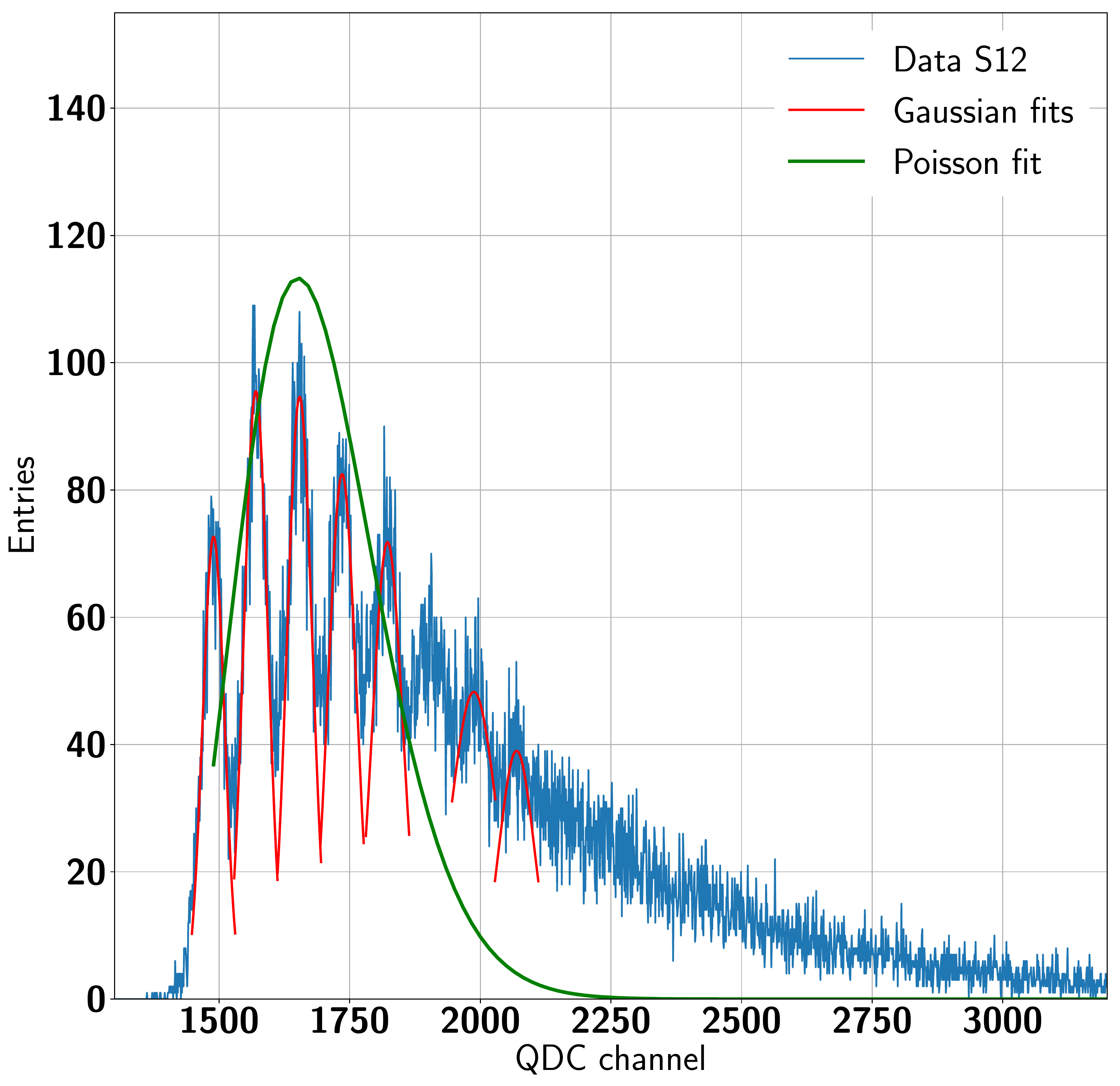}
\end{minipage}
\raggedleft
\begin{minipage}{0.49\linewidth}
	\includegraphics[width =\linewidth]{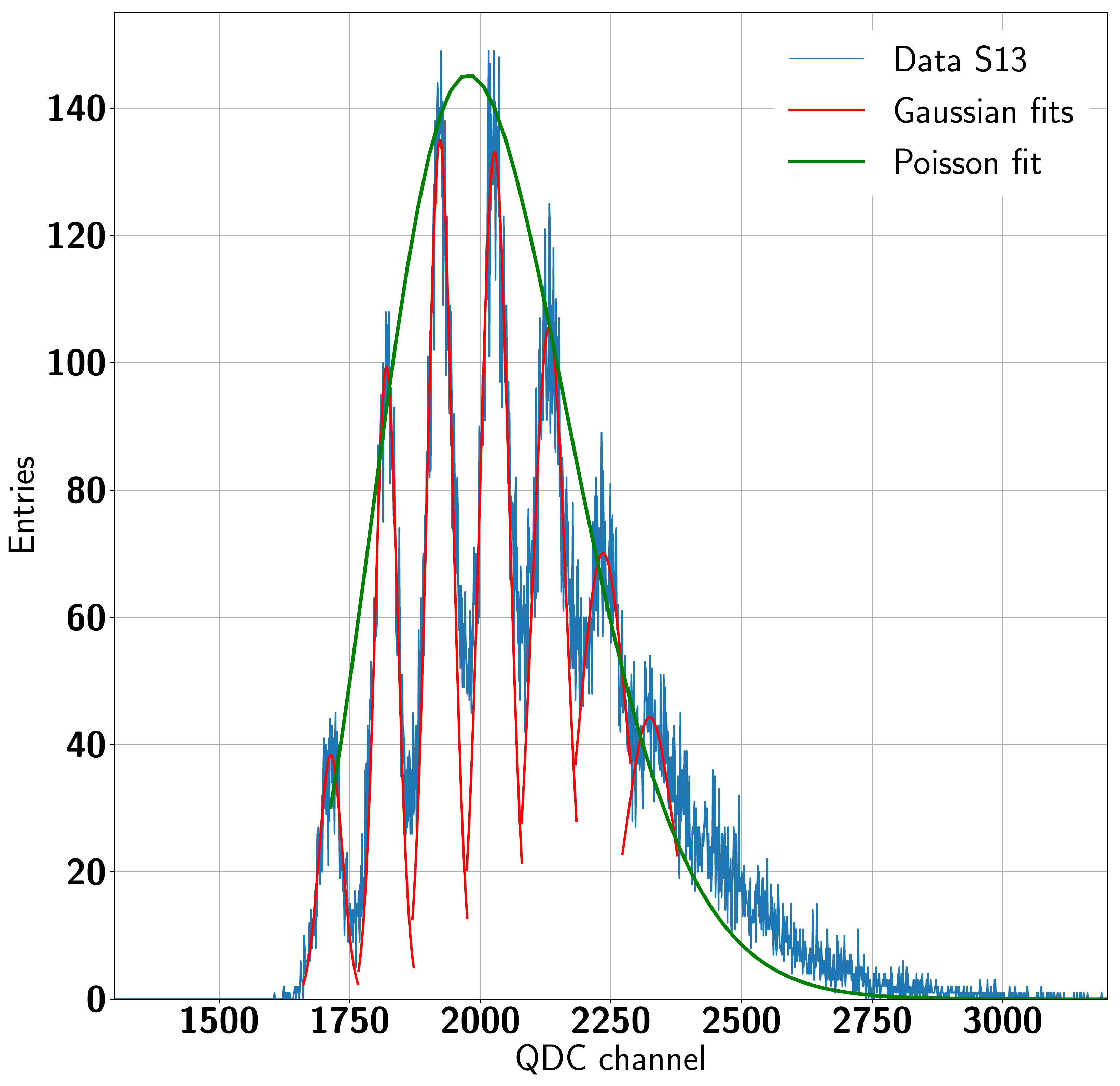}
\end{minipage}
\caption{Poisson fit of the charge spectrum for SiPM array \textit{S12} (right) and \textit{S13} (left). The number of incident photons was for both measurements about 6.8 photons per pulse.}
\label{pic:example_poisson_fit}
\end{figure}
Examples of the Poisson fit for two charge spectra of array \textit{S12} and \textit{S13} are shown in figure \ref{pic:example_poisson_fit}. Both measurements have been performed with a comparable number of incident photons (\textit{S12}: 6.77 photons/pulse, \textit{S13}: 6.83 photons/pulse). For array \textit{S13}, the Poisson fit resembles the form of the charge spectrum and gives a good estimate for the mean number of fired APDs. Regarding array \textit{S12}, the fit can be performed but is disturbed by the long tail of the charge spectrum coming from afterpulses, crosstalk and dark counts. Due to this, the fit result for the mean number of fired APDs is shifted towards higher values for an increasing number of incoming photons.\\
The number of firing APDs was calculated for several numbers of incident photons in the range of zero and 10 photons per pulse. The measurement has been performed for one channel of each array. For the SiPM array \textit{S13}, channel A3 (channel number 3) has been used. The measured channel of array \textit{S12} was B4 (channel number 12).\\
\begin{figure}[t]
	\centering
	\includegraphics[width=0.7\linewidth]{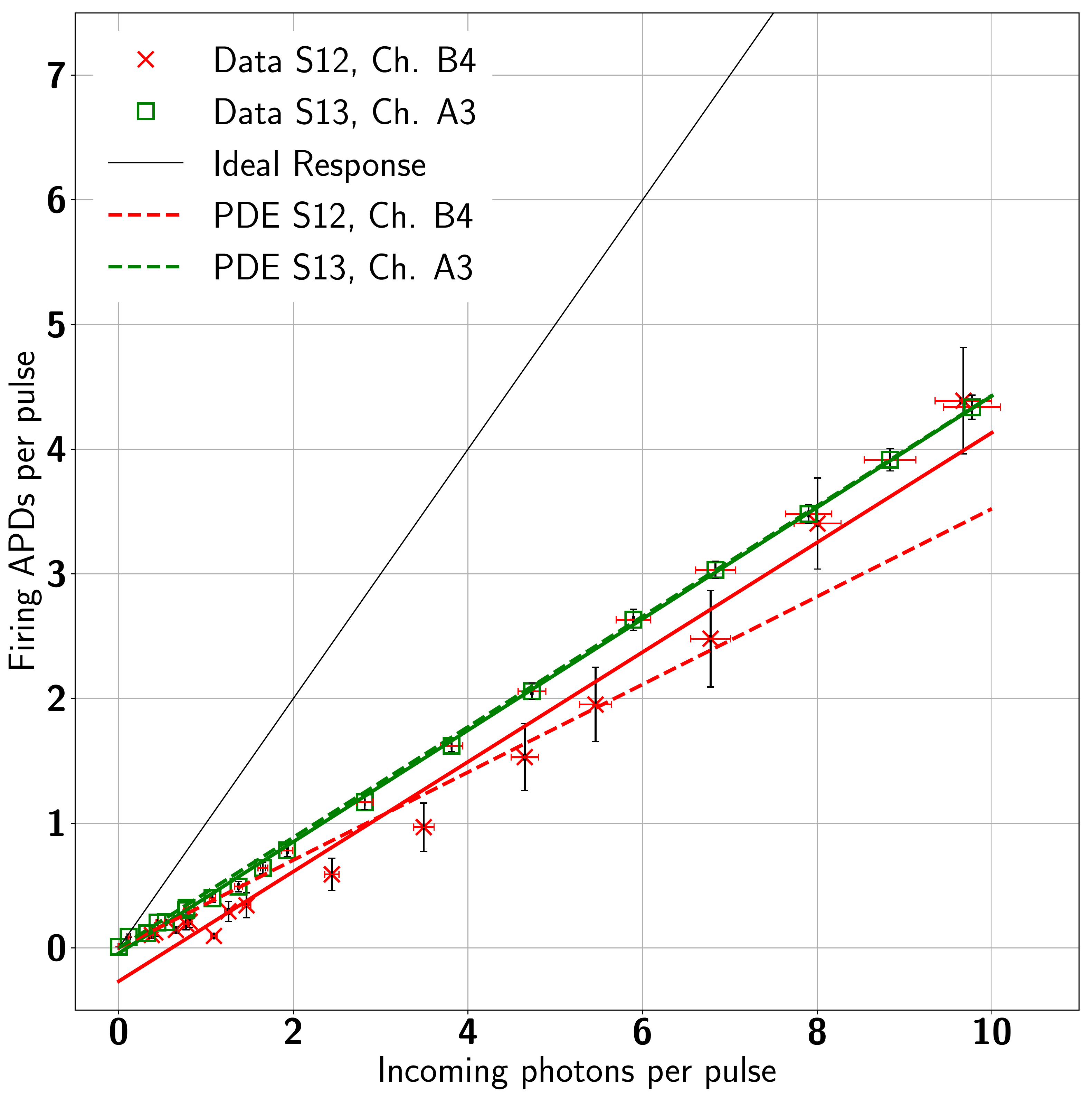}
	\caption{Response behaviour measurement of one channel of each SiPM array \textit{S13361-3050AS-08} 
	(green boxes) and \textit{S12642-0808PA-50} (red crosses). Statistical errors are shown in black, 
	systematic errors are shown in red. The ideal response is shown as a black line. 
	Fits of the SiPM response behaviours are included as coloured lines. The theoretical response 
	behaviour calculated with the measured PDEs is plotted with dotted coloured lines for each channel.
	}
	\label{pic:compare_response}
\end{figure}
Figure~\ref{pic:compare_response} shows the results of the response behaviour measurement. Linear functions are fitted to the slopes of the measured data. The fit results in
\begin{align}
S12642:&\hspace{5mm} N_{APD} = (0.440 \pm 0.019) \cdot N - (0.268 \pm 0.082) \\
S13361:&\hspace{5mm} N_{APD} = (0.447 \pm 0.003) \cdot N - (0.043 \pm 0.014).
\end{align}
The ideal SiPM response, meaning one firing APD per incoming photon is included in 
figure~\ref{pic:compare_response} as a black solid line. The theoretical response behaviour 
calculated with the measured PDE for each of the two channels is plotted with dotted lines. 
This response behaviour follows
\begin{align}
S12642, \text{Channel B4}:&\hspace{5mm} N_{APD} = 0.3522 \cdot N \\
S13361, \text{Channel A3}:&\hspace{5mm} N_{APD} = 0.4431 \cdot N.
\end{align}
The response behaviour of both SiPM arrays have lower inclinations than the ideal response line which is expected 
since both arrays have less than $100\%$ PDE. For array \textit{S13} the linear fit to the measured values and the theoretical behaviour calculated from the PDE measured in the characterization measurement agree. Regarding the results of array \textit{S12}, the measured values do not follow the predicted behaviour. This has its reason in the Poisson fit of the charge spectrum, which gives too high results for the mean number of firing APDs for an increasing numbers of incident photons (see fig. \ref{pic:compare_response}). Other possibilities for getting the mean number of firing APDs are to calculate the mean or the median of the charge spectrum and determine $N_{APD}$ out of this. But due to the long tail of the \textit{S12}-SiPM charge spectrum these methods lead to an even higher overestimation of $N_{APD}$.\\
Looking at the results for array \textit{S13}, a linear response behaviour of SiPMs for low numbers of incident photons is visible. The result for the PDE obtained from the characterization measurement and the response behaviour are in coincidence.\\ 
Due to the low number of incident photons per pulse, the examined region of the SiPM response is far away 
from the region in which the SiPM saturates. Measurements of the dynamic range up to saturation 
have been performed elsewhere, e.g. in~\cite{Bretz1}. 

\subsection{UV sensitivity}
\label{subsec:UV}
\begin{figure}
	\centering
	\includegraphics[width = \linewidth]{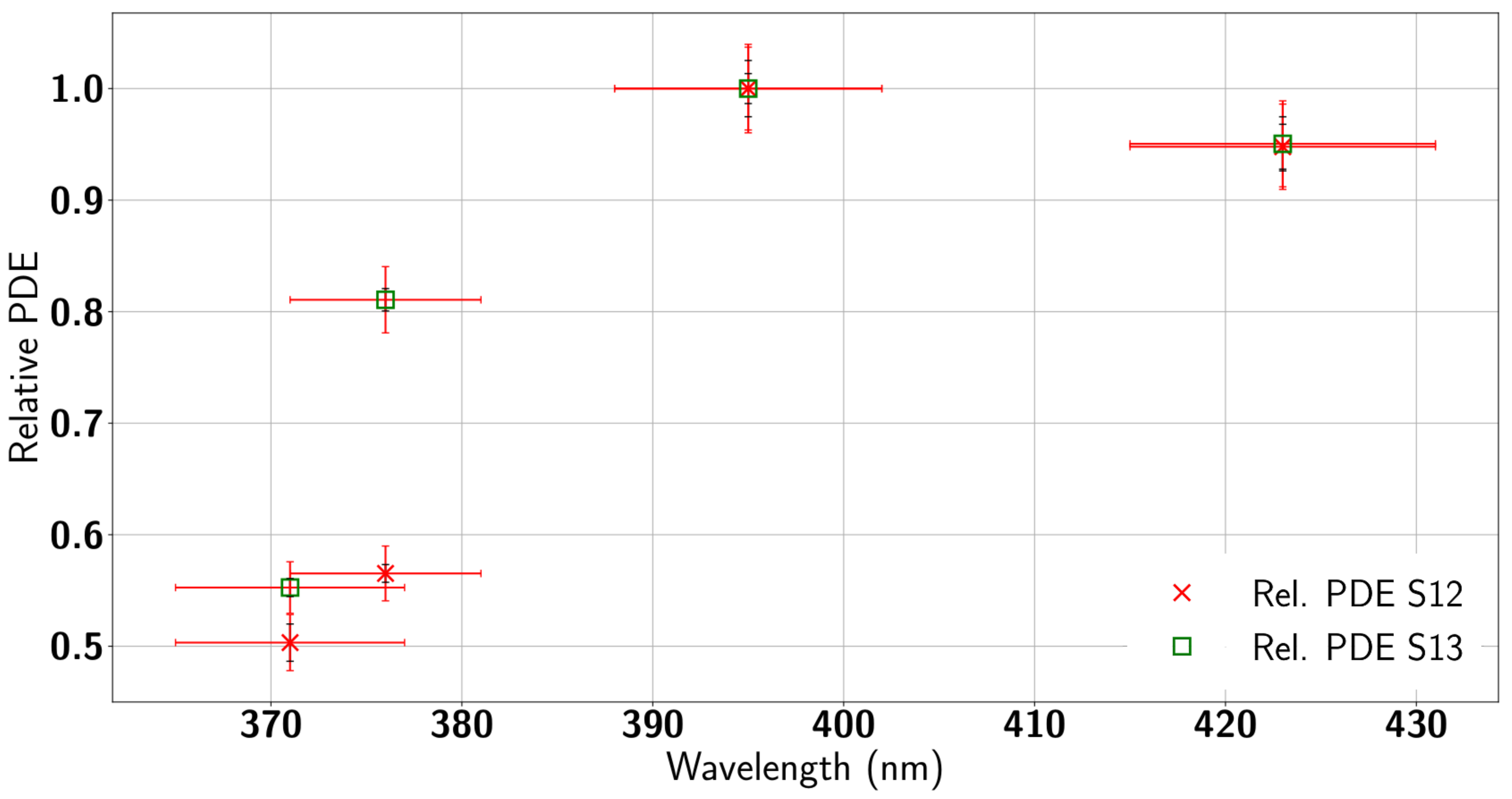}
	\caption{UV sensitivity measurement of one channel of each SiPM array \textit{S13361-3050AS-08} 
	(green boxes) and \textit{S12642-0808PA-50} (red crosses) at the wavelengths 
	$423\,$nm, $395\,$nm, $376\,$nm and $371\,$nm. 
	Statistical errors are shown in black, systematic errors are shown in red.}
	\label{pic:compare_UV}
\end{figure}

For comparing the UV sensitivity of both arrays, the PDE of one channel of each array was measured at 
different wavelengths. The available wavelengths are shown in table~\ref{tab:Setup-SPOCK_LEDarrays}. 
The measurement was done with the B4-channels of both arrays (channel number 12).\\
Figure~\ref{pic:compare_UV} shows the results of the measurement. 
The relative PDE shown on the y-axis in figure~\ref{pic:compare_UV} is the measured PDE normalized 
to the highest measured PDE value of each channel which was for both channels at a wavelength of $395\,$nm. 
The improvement of SiPM array \textit{S13} regarding the sensitivity in UV photons is clearly visible. 
For wavelengths lower than $395\,$nm, the relative PDE of array \textit{S13} is apparently higher which 
is due to the different resin of the arrays.\\
Regarding wavelengths below $371\,$nm, the manufacturer claims sensitivity of the SiPMs to wavelengths 
down to $320\,$nm for the epoxy resin array and $270\,$nm for the silicone resin array.

\subsection{Temperature dependent measurements}
\label{sec:temp}
\begin{figure}
	\centering
	\includegraphics[width = \linewidth]{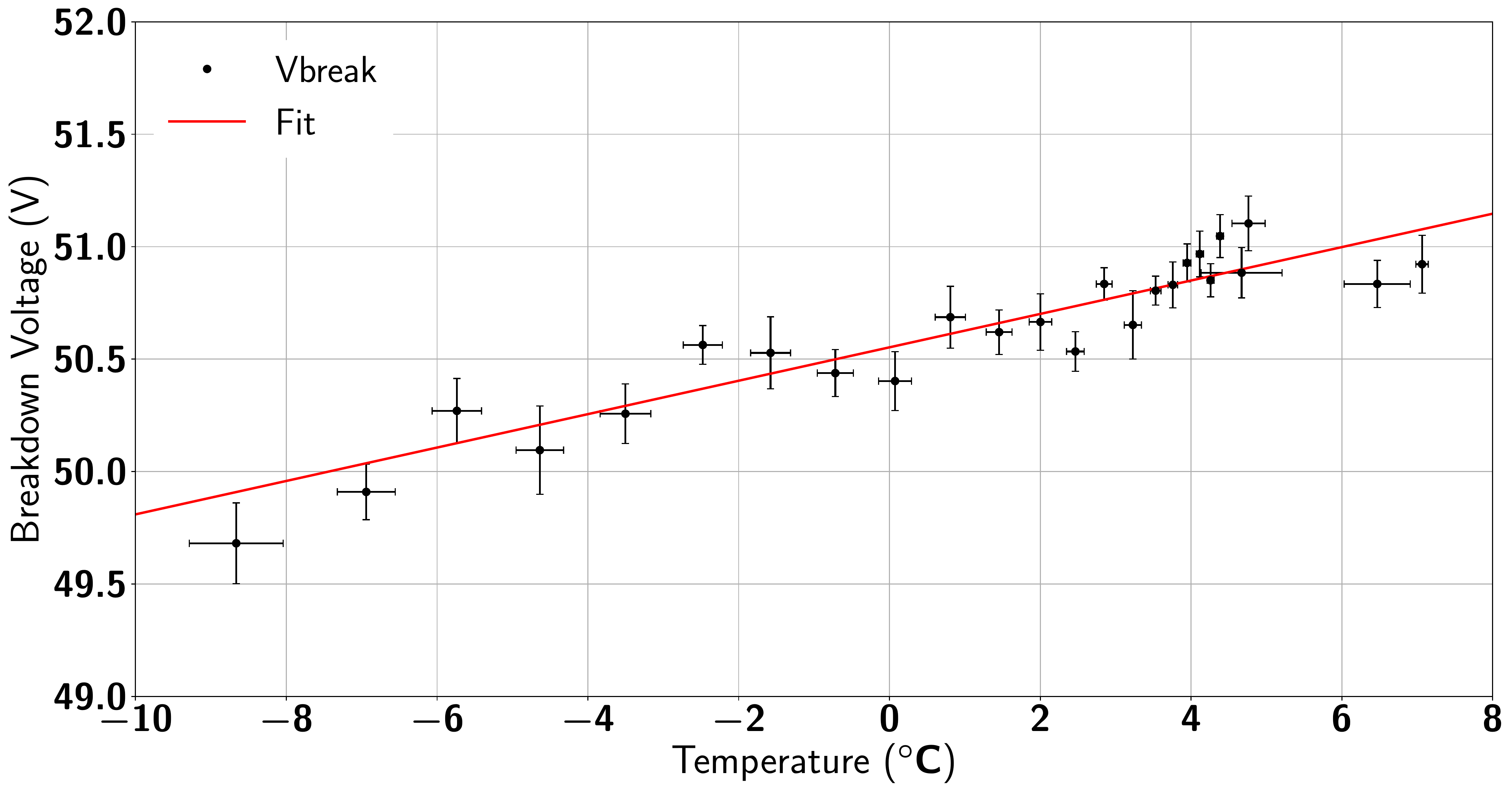}
	\caption{Temperature dependent measurement of the breakdown voltage. 
	The uncertainty shown is coming from the uncertainty in the fit parameters.}
	\label{pic:temp_Vbreaks}
\end{figure}

\begin{figure}
	\centering
	\includegraphics[width = \linewidth]{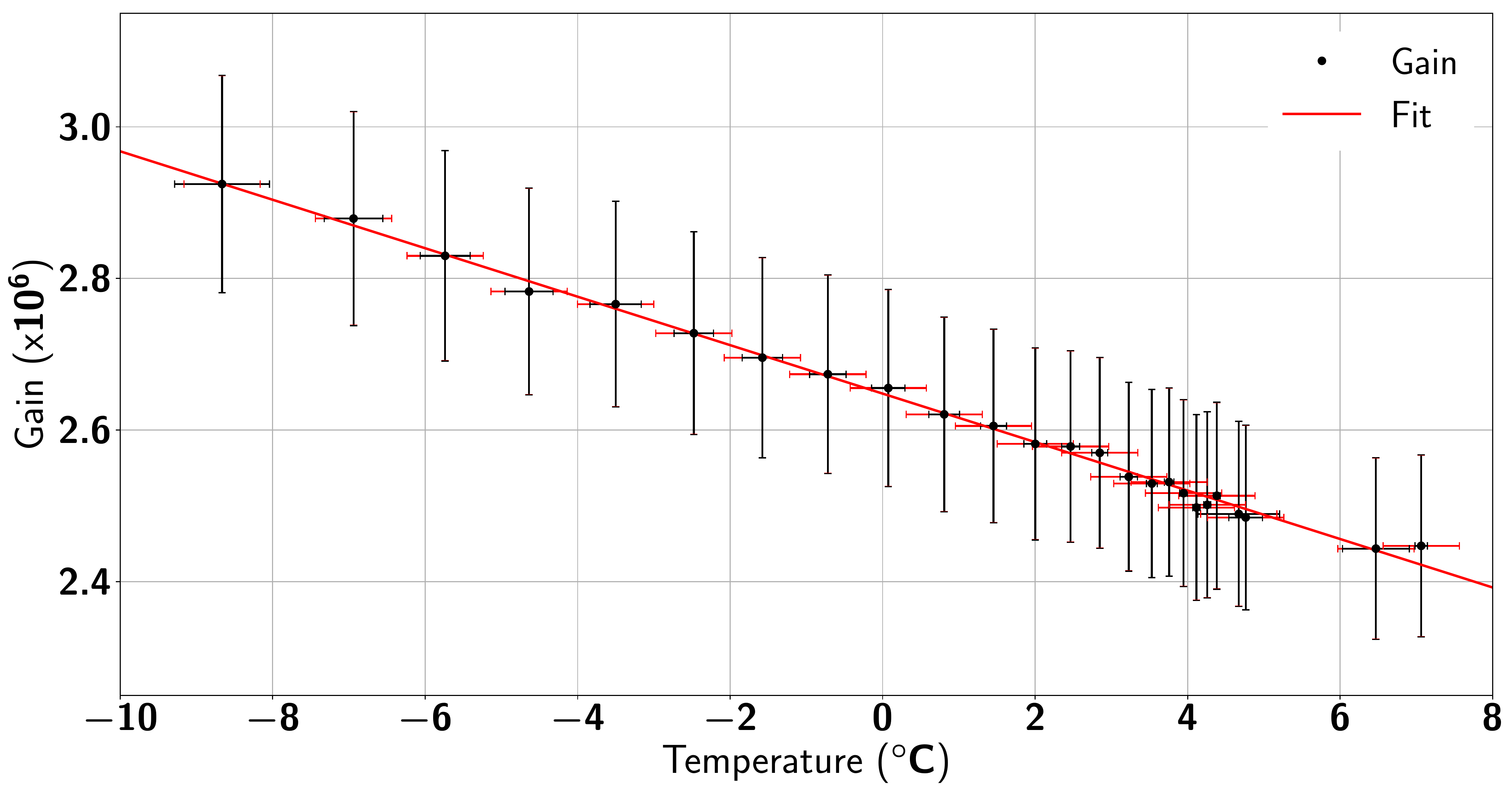}
	\caption{Temperature dependent measurement of the gain. The bias voltage was kept constant at $55.2\,$V. Statistical errors are shown in black, systematic errors are shown in red. }
	\label{pic:temp_Gain}
\end{figure}

\begin{figure}
	\centering
	\includegraphics[width = \linewidth]{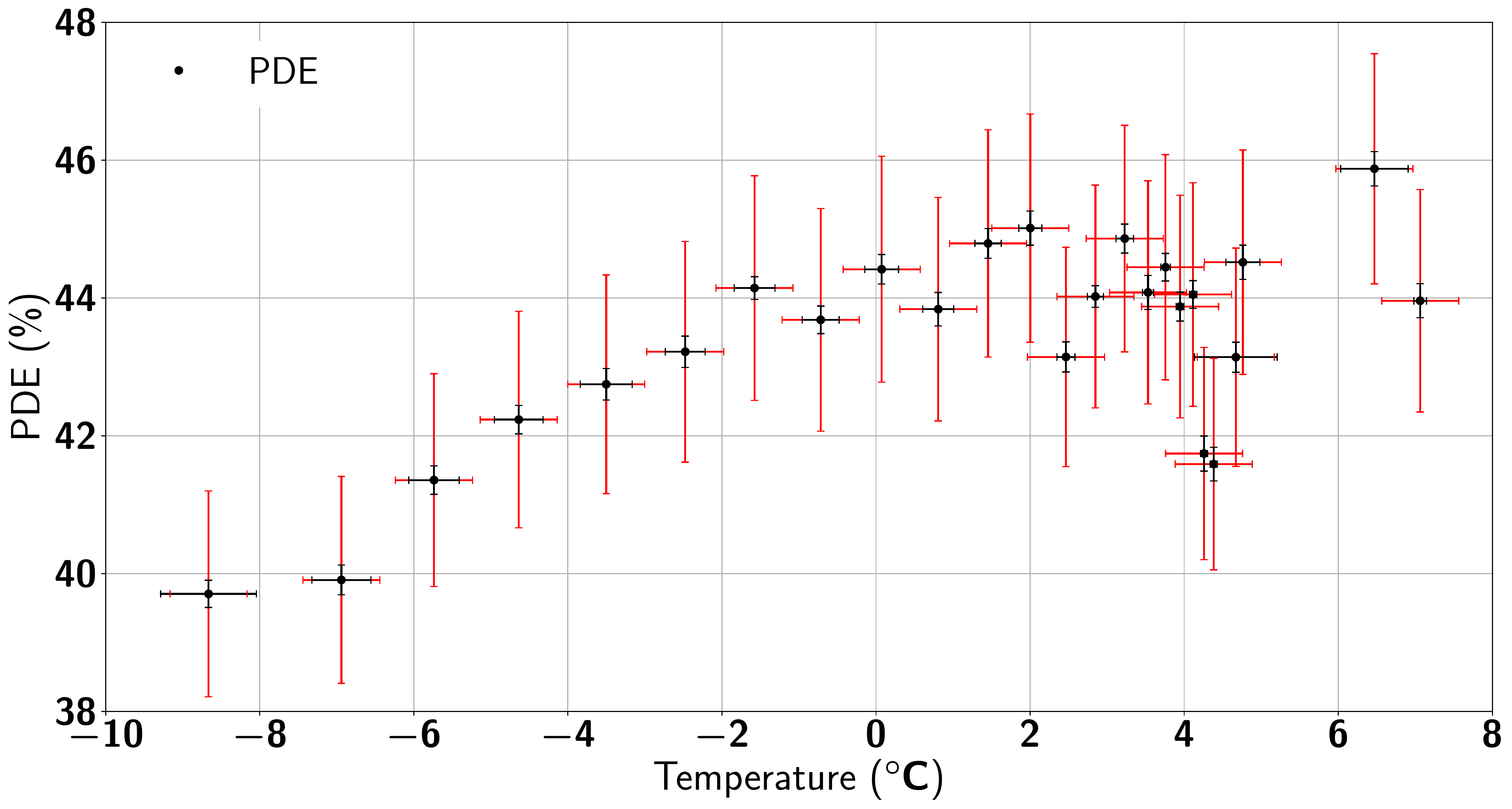}
	\caption{Temperature dependent measurement of the PDE. The bias voltage was kept constant 
	at $55.2\,$V. The incident photons hat a wavelength of $(423\pm8)\,$nm. 
	Statistical errors are shown in black, systematic errors are shown in red.}
	\label{pic:temp_PDE}
\end{figure}

\begin{figure}
	\includegraphics[width = \linewidth]{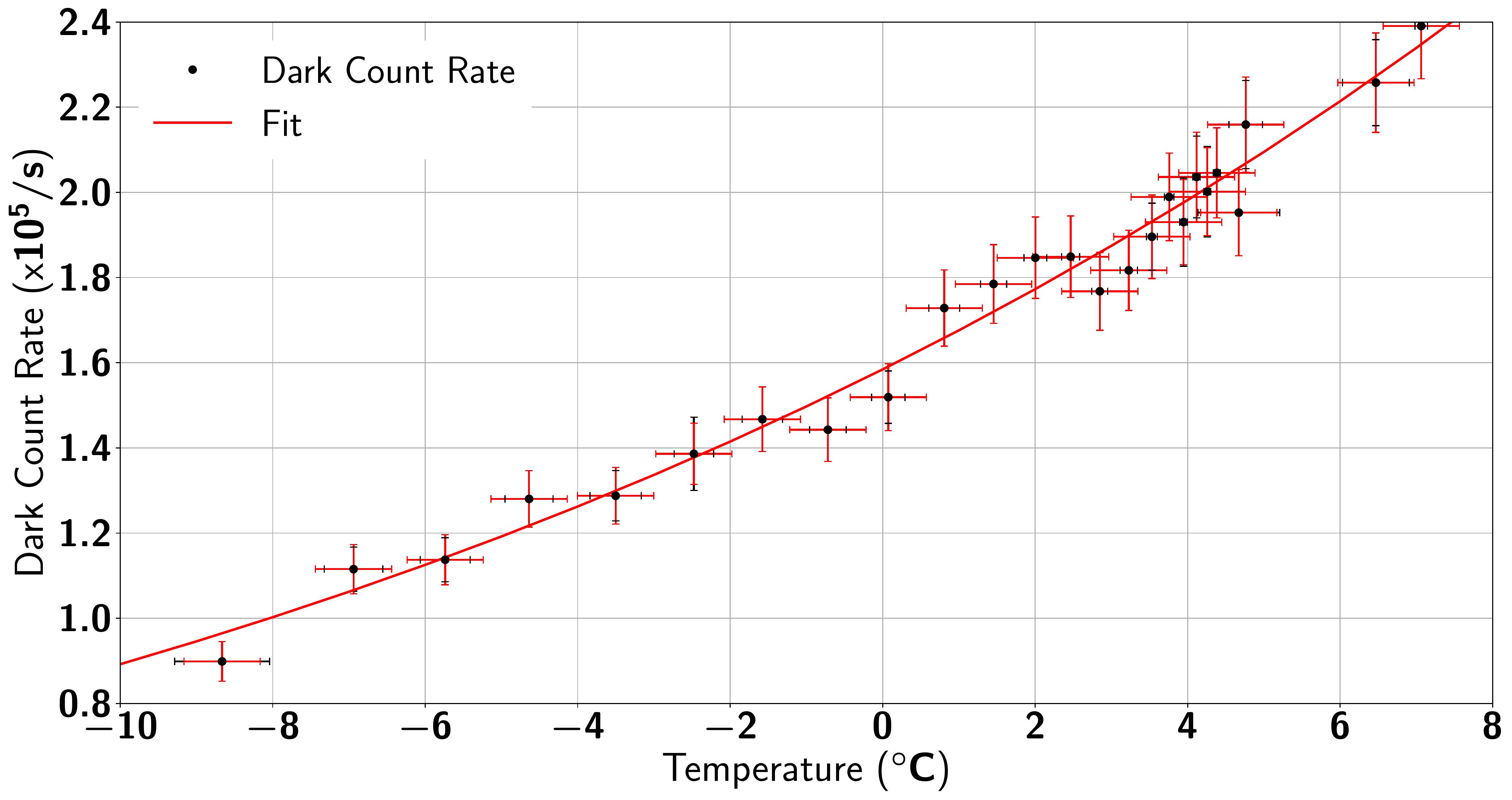}
	\caption{Temperature dependent measurement of the dark-count rate at a bias voltage 
	of $55.2\,$V. Statistical errors are shown in black, systematic errors are shown in red. }
	\label{pic:temp_DC}
\end{figure}

Temperature dependent measurements have been performed with channel E4 (channel number 36) of SiPM 
array \textit{S13361-3050AS-08}. The SiPM was put into a cooling box together with thermal reservoirs. 
The cooling box with the SiPM placed inside was cooled down in a fridge to about $-10^{\circ}$C. The 
measurements were performed as the cooling chamber warmed up. For the measurements the cooling 
chamber was placed inside of the photon shielding of SPOCK. The temperature rose from $-10^{\circ}$C to $8^{
\circ}$C and was constantly monitored with a \textit{DS18B20} temperature sensor placed near the SiPM. 
The bias voltage was kept constant at $55.2\,$V.\\
\paragraph{Breakdown voltage} The temperature dependence of the breakdown voltage is shown in 
figure~\ref{pic:temp_Vbreaks}. The breakdown voltage increases linearly with increasing temperature 
which is the expected behaviour~\cite{SiPM_Char2, HamaHandbook, SiPM_Char3}. 
The result for the fit of the breakdown voltage values shown in figure \ref{pic:temp_Vbreaks} is
\begin{equation}
V_{break}(T) = (0.074 \pm 0.007) \frac{\textrm{V}}{^\circ \textrm{C}} \cdot T + (50.552 \pm 0.029)\, \textrm{V} \, .
\end{equation}
The temperature coefficient which gives information about the breakdown voltage behaviour 
regarding the temperature is measured as
\begin{equation}
\Delta V_{Break,\textit{S13}} = (74 \pm 7) \frac{\textrm{mV}}{^\circ \textrm{C}}.
\end{equation}
The temperature coefficient provided by Hamamatsu as 
$\Delta V_{Break,\textit{S13}} = 60\, \textrm{mV} / ^\circ \textrm{C} $ for the SiPM 
array \textit{S13} is off the measured value.
\paragraph{Gain} In figure \ref{pic:temp_Gain} the measured temperature dependency of the gain is shown. 
Since the bias voltage was kept constant and the breakdown voltage increased with increasing 
temperature, the gain is expected to decrease. 
The measurement confirms this behaviour. The gain decreases by about 10\% per 10 degree in the measured temperature range. 
The fit results are 
\begin{equation}
G(T) = -0.032 \frac{10^6}{^\circ \textrm{C}} \cdot T + (2.648 \pm 0.002) \cdot 10^6.
\label{equ:Measurements-gain_temperature}
\end{equation}
\paragraph{PDE} A statement about the PDE temperature dependency can not be made. 
Since the SiPM was biased with a constant voltage, the PDE should decrease with increasing temperature 
due to the increase of the breakdown voltage.  The results of the measurements are shown in 
figure \ref{pic:temp_PDE}, where it is seen that the PDE values do not follow this expectation.  
For temperatures below the freezing point the PDE rises. Above the freezing point, the PDE values seem 
to be more or less constant.\\
This rather strange behaviour of the PDE could be caused by the formation of an ice or frost layer on 
the SiPM's surface below $0^{\circ}$C preventing photons to penetrate the SiPM. 
The SiPM was cooled down inside the cooling chamber together with the thermal reservoirs. 
Although the SiPM surface was cleaned from ice before the measurement, a new layer might 
have accumulated before the first measurements. With increasing temperature, the ice 
layer became thinner which results in a higher PDE. Above the freezing point the ice melted and 
water might have covered parts of the SiPM surface leading to the more random like distribution 
for temperatures higher than zero degrees. Since the PDE is the only measured SiPM characteristic 
that is sensitive to changes in the number of penetrating photons during the measurement, 
the other characteristics are not influenced by this circumstance.\\
\paragraph{Dark-count rate} Figure \ref{pic:temp_DC} shows the dark-count rate measurement results. 
An increase in dark-count rate with an increase in temperature is clearly visible. 
Since dark counts are thermally excited electrons, an increase of the dark-count rate with temperature is 
expected. For a constant gain, the dark-count rate $\xi(T)$ should follow
\begin{equation}
\xi(T) = A \cdot T^{3/2} \cdot e^{\frac{E_g}{2 \cdot k_B \cdot T}}.
\label{equation1}
\end{equation}
with the absolute temperature $T$, the band gap energy $E_g$, the Boltzman constant $k_B$ and an arbitrary constant $A$~\cite{HamaHandbook}. \\
In this measurement, the bias voltage of the SiPM was kept constant and the gain decreased 
with increasing temperature (see fig. \ref{pic:temp_Gain}). Therefore, equation \ref{equation1} 
is not valid. In the measured region the fit
\begin{equation}
\xi(T) = (4.55 \pm 0.03) \cdot 10^{-33} \textrm{Hz} \cdot T (K)^{15.4}
\label{equ:Measurements-dc_fit}
\end{equation}
gives the best results.

\section{Conclusion}

The characterization-measurement results including breakdown voltage, gain, photon detection efficiency, 
dark-count rate and crosstalk probability of the two SiPM arrays \textit{S12642-0808PA-50} and 
\textit{S13361-3050AS-08} confirm to a large extent the SiPM characteristics given by the manufacturer. 
The latter SiPM array improved in all of the measured characteristics compared to the older series 
SiPM array (see table~\ref{tab:measurements-comp.results} for details). 
Additional measurements of the response behaviour and of the UV sensitivity show a linear response 
behaviour for small numbers of incident photons of both arrays and a better UV sensitivity for 
wavelengths below $395\,$nm of the array \textit{S13361-3050AS-08}. 
Measurements performed with one channel of array \textit{S13361-3050AS-08} 
confirm the strong temperature dependence of the SiPM characteristics.

Due to its better performance, the SiPM array \textit{S13361-3050AS-08} has been chosen to be employed 
in SiECA\footnote{Silicon Elementary Cell Add-on}. The uniformity of the array is sufficient to build a 
focal surface out of SiPM arrays. Inhomogeneities between different channels of the SiPM array can be 
handled by applying individual bias voltages to every channel like it is done in the operation version of 
SiECA. 

The presented newly installed measurement setup at KIT (SPOCK) and the applied methods allow for 
efficient tests and characterization of SiPMs and SiPM arrays of all kinds and geometries. 
I.e., the precise characterization of SiPMs can be done for manifold applications and future experiments.

\section*{Acknowledgements}
We kindly acknowledge the support by the 'Helmholtz Alliance for Astroparticle Physics - HAP' funded by 
the Initiative and Networking Fund of the Helmholtz Association. 
Acknowledgements go also to Hamamatsu Photonics for giving us the opportunity to work with their 
newest SiPM products. 
The authors deeply acknowledge all the support by the JEM-EUSO collaboration and in particular by the 
EUSO-SPB team. The author wants to thank expressly Lawrence Wiencke for his advice during the 
writing process. An important contribution to the success of this work was made by the technical 
staff of the IKP at KIT, in particular Heike Bolz and Bernd Hoffmann, as well as by 
the students Sally-Ann Sandkuhl, Simon Ehnle, Nils Hampe and Sonja Schneidewind.

\bibliography{mybibfile}

\end{document}